\documentclass[twocolumn,english,showpacs,footinbib,bibnotes]{revtex4-1}

\usepackage[version=3]{mhchem}

\usepackage[usenames,dvipsnames]{color}
\usepackage{graphicx}
\usepackage{dcolumn}
\usepackage{bbm}
\usepackage{bm}
\usepackage{amsmath}
\usepackage{amssymb}
\usepackage{times}
\usepackage{placeins}
\usepackage{hyperref}
\hypersetup{colorlinks=false}

\usepackage{multirow}

\newcommand{\mr}[1]{\multirow{2}{*}{#1}}

\newcommand{\ie}{{i.e.~}}

\graphicspath{{Figures/}}

\begin{document}

\title{Classical Spin Spirals in Frustrated Magnets from Free-Fermion Band Topology}

\author{Jan Attig}
\author{Simon Trebst}
\affiliation{Institute for Theoretical Physics, University of Cologne, 50937 Cologne, Germany}

\date{\today}

\begin{abstract}
The formation of coplanar spin spirals is a common motif in the magnetic ordering of many frustrated magnets.
For classical antiferromagnets, geometric frustration can lead to a massively degenerate ground state manifold of spirals whose propagation vectors can be described, depending on the lattice geometry, by points (triangular), lines (fcc), surfaces (frustrated diamond) or completely flat bands (pyrochlore).
Here we demonstrate an exact mathematical correspondence of these spiral manifolds of classical antiferromagnets with the Fermi surfaces of free-fermion band structures. We provide an explicit lattice construction relating the frustrated spin model to a corresponding free-fermion tight-binding model. Examples of this correspondence relate the 120$^\circ$ order of the triangular lattice anti\-ferromagnet to the Dirac nodal structure of the honeycomb tight-binding model or the spiral line manifold of the fcc antiferromagnet to the Dirac nodal line of the diamond tight-binding model. 
We discuss implications of topological band structures in the fermionic system to the corresponding classical spin system.
\end{abstract}

\maketitle


\section{Introduction}

In frustrated magnets, competing interactions 
give rise to many nearly degenerate low-energy states and a characteristic residual entropy down to zero temperature \cite{Ramirez1994}.
In the presence of such an abundance of low-energy states, these magnets oftentimes evade thermal ordering around the Curie-Weiss temperature as the system remains fluctuating within the manifold of nearly degenerate states down to considerably lower temperature scales -- giving rise to a regime that is commonly referred to as cooperative paramagnet or spin liquid \cite{Balents2010}.
The emergence of such a massive degeneracy of distinct ground states is sometimes coined ``accidental'' as there is no distinct symmetry mechanism protecting it or instigating it in the first place. Nevertheless, for specific systems one can find elegant ways to describe the physics of such highly degenerate ground state manifolds. This includes the emergence of Coulomb phases \cite{Henley2010} in the presence of local constraints (as, for instance, in spin ice \cite{Castelnovo2008}) or the formation of spiral surfaces \cite{Smart1966,Bergman2007,Reimers1991,Moessner1998} in a broad family of geometrically frustrated Heisenberg antiferromagnets, which will be of interest in this manuscript.
 
\begin{figure}[b]
  \centering
  \includegraphics[width=\linewidth]{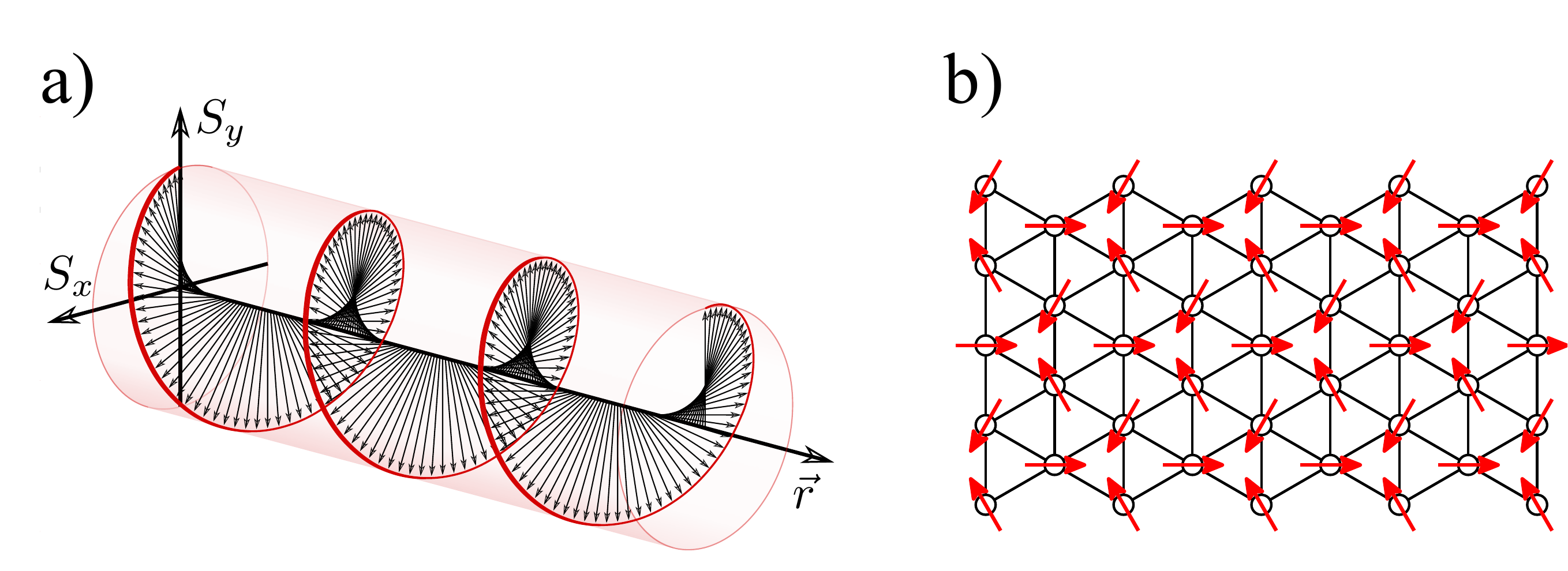}
  \caption{ {\bf Coplanar spirals}.
  		(a) A coplanar spin spiral can be described by a single propagation vector $\vec{q}$ in reciprocal space
  		     indicating its direction and pitch.
		(b) The 120$^\circ$ order of the triangular lattice antiferromagnet is a familiar example of a spin spiral. 
		      The two possible orientations correspond to propagation vectors 
		      $\vec{q} = \left({2\pi}/{\sqrt{3}}, \pm {2\pi}/{3}\right)$,
		      respectively.
		}
  \label{fig:SpinSpirals}
\end{figure}

\begin{figure}
  \centering
  \includegraphics[width=\linewidth]{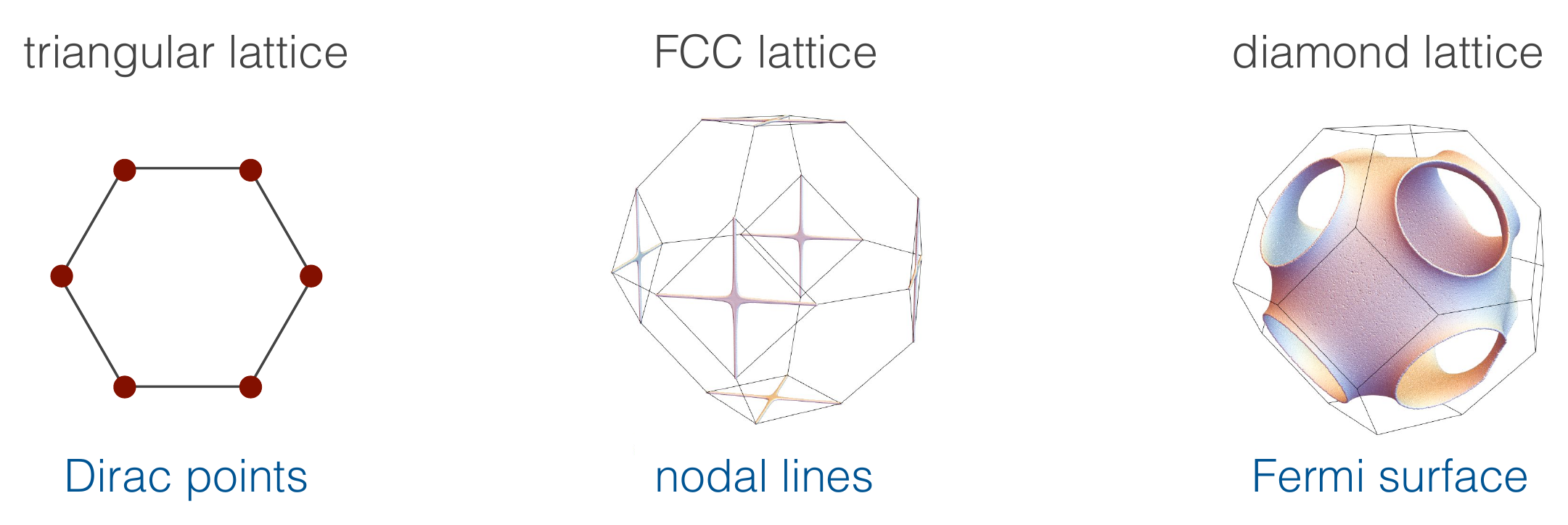}
  \caption{{\bf Spiral manifolds}. Geometrically frustrated classical antiferromagnets 
  		on the triangular, fcc, and frustrated diamond lattice
		exhibit degenerate spiral ground states whose propagation vectors form distinct
		manifolds in reciprocal space. 
  	        The variety of manifolds is strikingly reminiscent of nodal structure / Fermi surfaces 
	        of electronic systems as indicated in the bottom line.
		}
  \label{fig:SpiralManifolds}
\end{figure}

For classical Heisenberg antiferromagnets it has long been appreciated that competing interactions generically lead to the formation of coplanar spin spirals \cite{Kaplan1959,Yoshimori1959,Villain1959}. Indeed, spin spirals are an ubiquitous motif in the magnetic ordering of many frustrated magnets \cite{Kaplan2007}. With a single coplanar spiral being uniquely described by a propagation vector (indicating its direction and pitch),
one can express sets of degenerate spiral states by the manifold of their respective propagation vectors. For instance, the two possible orientations of the 120$^\circ$ order of the triangular lattice anti\-ferromagnet, illustrated in Fig.~\ref{fig:SpinSpirals}, are captured by the two propagation vectors $\vec{q} = \left( {2\pi}/{\sqrt{3}}, \pm {2\pi}/{3}\right)$. The multitude of degenerate ground states of the fcc antiferromagnet can be described by all $q$-vectors along a {\em line} in reciprocal space \cite{Smart1966}. Even higher degeneracies are encountered for frustrated diamond lattice antiferromagnets (with both nearest and next-nearest neighbor couplings) where the spiral propagation vectors form a two-dimensional {\em surface} \cite{Bergman2007}, as illustrated in Fig.~\ref{fig:SpiralManifolds}, and the pyrochlore antiferromagnet where any propagation vector in the {\em volume} of the Brillouin zone is permissible \cite{Reimers1991,Moessner1998}, thus indicating an extensive degeneracy.

Given this variety of distinct spiral manifolds in reciprocal space what inevitably comes to mind is a striking resemblance  
to Fermi surfaces of electronic systems. While in most metals the conductance and valence bands touch along a two-dimensional Fermi surface, 
some semi-metals exhibit band touchings confined to nodal lines or even single points, as in the case of Dirac or Weyl semi-metals \cite{Armitage2017}. 
It is thus natural to ask whether one can make a one-to-one correspondence between spiral manifolds and Fermi surfaces. However, it is far from obvious that one can establish any correspondence at all, since one is dealing with a manifold of {\em minimal-energy} ground states in a {\em classical} system on the one hand, and with a manifold of {\em mid-energy} states (at the Fermi level) in a {\em quantum} system on the other hand. In addition, the classical and quantum systems in such a correspondence must have the {\em same} spatial dimensionality, which is adverse to the often employed correspondence between a quantum system in $d$ spatial dimensions and a classical system in $d+1$ spatial dimensions.

It is the purpose of this manuscript to describe exactly such a correspondence between spiral manifolds in frustrated classical antiferromagnets and Fermi surfaces in electronic systems captured by simple free-fermion tight-binding models. 
The basic idea is that by {\em squaring} the Hamiltonian of the fermionic system one can move the states at the Fermi level ($E_F=0$) of a tight-binding Hamiltonian (with symmetric energy spectrum) to turn into its minimal eigenvalues.
Using precisely this idea, one can then formulate a general matrix identity 
\begin{equation}
	{\bf M}(\vec{k}) = {\bf H}(\vec{k})^2 - E_0 \cdot {\bf 1} 
	\label{eq:MatrixCorrespondence}
\end{equation}
that relates the Fourier transformed spin-coupling matrix ${\bf M}(\vec{k})$ and its minimal eigenvalues (plus associated eigenvectors)  
to the tight-binding matrix ${\bf H}(\vec{k})$ of a corresponding free fermion system and its eigenvalues (plus associated eigenvectors). 
The additional offset $E_0$ is precisely the ground-state energy of the classical spin system. 
The above matrix correspondence 
thereby allows to establish a rigorous connection between the highly degenerate ground states of the spin system and nodal features of the fermionic band structure and pinpoints the sought-after identification of spiral surfaces and Fermi surfaces. 
In addition, we provide an explicit lattice construction that allows to connect specific spin and fermion lattice models on the left and right hand side of the matrix correspondence \eqref{eq:MatrixCorrespondence}.
Specifically, this lattice construction gives an explicit meaning to the ``squaring'' of the fermionic Hamiltonian -- it is the sublattice of next-nearest neighbor sites in the fermion lattice that can be identified with the respective spin lattice.
Vice versa, going from the spin model to the fermion model (\ie taking a ``square root'') is achieved by systematically replacing fully-connected plaquettes (such as triangles) of the spin lattice by individual sites of the corresponding fermion lattice. 
Examples of this construction relate, for instance, the 120$^\circ$ order of the triangular lattice anti\-ferromagnet to the Dirac nodal structure of the honeycomb tight-binding model or the spiral line manifold of the fcc antiferromagnet to the Dirac nodal line of the diamond tight-binding model. A number of further examples of this correspondence are summarized in Table \ref{tab:overview}.

The matrix correspondence \eqref{eq:MatrixCorrespondence} and its associated lattice construction allow to reveal a number of additional connections. For instance, one can ask how topological aspects of the fermionic band structure play out in the classical magnet -- a perspective which we will discuss along general symmetry arguments by considering the free-fermion models of \eqref{eq:MatrixCorrespondence} in terms of the 
classification of topological insulators \cite{Schnyder2008,Kitaev2009,Ryu2010} rooted in the symmetry classification of free-fermion systems \cite{Altland1997}. We will also consider more specific situations and discuss the occurrence and absence of edge states or the effect of strain in triangular lattice antiferromagnets. Vice versa, identifying spin systems with extensive ground-state degeneracies allows to engineer fermionic band structures with completely flat bands -- an essential ingredient, for instance, for models of interacting electrons  exhibiting fractional Chern insulators \cite{Sheng2011,Tang2011,Sun2011,Neupert2011,Regnault2011}.

Before we indulge in a detailed discussion of our results in the remainder of the manuscript, we note that identifying such a general matrix correspondence \eqref{eq:MatrixCorrespondence} between a classical system and free-fermion quantum system (of identical spatial dimensionality) is reminiscent of the recent work of Kane and Lubensky \cite{KaneLubensky2014} introducing the concept of ``topological mechanics'', which is based on a similar matrix correspondence between the classical Hamiltonian describing zero-frequency ``floppy modes'' of an isostatic lattice and a free-fermion Hamiltonian of the same spatial dimensionality.
Our work also connects to the earlier identification of parallels of classical normal modes in disordered systems and disordered, non-interacting fermion systems by Gurarie and Chalker \cite{Gurarie2002,Gurarie2003}.

\begin{table}[t]
	\begin{tabular}{c|c||c||c|c}
		\multicolumn{2}{c||}{\bf Heisenberg spins} &  & \multicolumn{2}{c}{\bf free fermions}  \\
		\mr{lattice} & \mr{LT} &  nodal  & \mr{lattice} & symmetry \\
		& &  structure  & & class \\[1mm]
		\hline \hline
		& & & &  \\[-2mm]
		triangular 	& \checkmark 	& points	& honeycomb 	& \multirow{2}{*}{BDI} \\
		FCC 		& \checkmark	& lines	& diamond	&  \\[1mm]
		\hline 
		& & & &  \\[-2mm]
		honeycomb $J_1$-$J_2$	& \checkmark 	& line & bi-honeycomb 	& \multirow{3}{*}{BDI} \\
		diamond $J_1$-$J_2$		 &  \checkmark	& surface & bi-diamond		&   \\
		bcc $J_1$-$J^*_2$	& 	\checkmark 	&  surface & bi-bcc			&   \\[1mm]
		\hline 
		& & & &  \\[-2mm]
		kagome & \checkmark & {\quad flat band \quad} & honeycomb-x &  \multirow{2}{*}{BDI} \\
		pyrochlore  & 	\checkmark & 	{\quad flat band \quad}	 	 & diamond-x &   \\[1mm]
	\end{tabular}
	\caption{{\bf Overview of results}.
		     The matrix correspondence \eqref{eq:MatrixCorrespondence} relates frustrated Heisenberg antiferromagnets
		     on the lattice given in the first column to a free-fermion tight-binding model on the lattice given in the fourth column.
		     In the central (third) column we provide the nodal structure that simultaneously describes the spiral manifold of the
		     spin model and the Fermi surface of the electronic model.
		     The second column indicates that the Luttinger-Tisza (LT) approximation \cite{Luttinger1951,Luttinger1946} is fully valid 
		     for all spin models considered in this table.
		     The fifth column indicates the symmetry class of the tight-binding model in the 10-fold way classification of free-fermion
		     models \cite{Altland1997}.
		     The lattice descriptor bi-honeycomb refers to a bilayer honeycomb lattice. 
		     The lattice descriptor honeycomb-x denotes an extended honeycomb
		     lattice as illustrated in Fig.~\ref{fig:kagome} below.
		     }
	\label{tab:overview}
\end{table}

The remainder of this manuscript is organized as follows. 
We will derive the spin fermion correspondence in Section \ref{sec:SpinFermionCorrespondence}. This includes a brief review
of the Luttinger-Tisza approximation for classical spin models and the tight-binding calculation for free-fermion models in Section \ref{sec:LT}
that motivates in rather general terms the matrix correspondence of Eq.~\eqref{eq:MatrixCorrespondence}. 
Section \ref{sec:Construction} is devoted to an explicit lattice construction relating the spin and free-fermion models summarized in Table \ref{tab:overview} above. We close Section \ref{sec:SpinFermionCorrespondence} with
a discussion of the general symmetry properties of the fermionic models in terms of the 10-fold way symmetry class classification.
Multiple case studies of the spin fermion correspondence are discussed in Section \ref{sec:CaseStudies} including magnetic Bravais and non-Bravais lattices as well as $J_1$-$J_2$ Heisenberg models on a variety of underlying lattices. 
Finally, in Section \ref{sec:topo} we discuss aspects of topological band structures in the free-fermion models with regard to the corresponding spin models. We close with a discussion in Section \ref{sec:discussion} and round off the manuscript with an appendix providing a short discussion of the validity/breakdown of the Luttinger-Tisza approximation.


\section{Spin Fermion Correspondence}
\label{sec:SpinFermionCorrespondence}

To formally derive the spin-fermion correspondence in this Section we proceed in multiple steps. 
We start by pointing out a number of analogies between the Luttinger-Tisza method to identify spiral ground states 
in classical Heisenberg spin models and tight-binding calculations for free-fermion systems. These observations naturally 
lead to the identification of the matrix correspondence \eqref{eq:MatrixCorrespondence}. We then proceed to discuss
a lattice correspondence that allows to explicitly construct a classical spin model from a free-fermion model and vice versa.
We conclude this formal part with a discussion of the symmetries of the free-fermion Hamiltonians considered in this 
spin-fermion correspondence and their classification in terms of the 10-fold way symmetry class classification.


\subsection{Luttinger-Tisza versus tight-binding calculations}
\label{sec:LT}

Our spin-fermion correspondence is rooted in an analogy of the Luttinger-Tisza approach to classical Heisenberg models
and tight-binding calculations of free-fermion systems. To set the stage, we start by shortly recapitulating the main steps
in both approaches.

\subsubsection*{Luttinger-Tisza method for classical spin systems}

The Luttinger-Tisza method \cite{Luttinger1951,Luttinger1946} is a rather general approach to identify the ground
states of a classical Heisenberg model. 
Its principal idea is to soften the constraint that all spins must have equal length while minimizing the energy,
and to identify the true ground state(s) of the spin system by those minimal eigenstate(s) that
meet the original hard spin constraint. 
It is this softening of the spin constraint that allows to find the energy minimum in a straightforward manner via a 
diagonalization of the interaction matrix in momentum space.

Let us discuss this approach by starting from a generic Heisenberg model
\begin{equation}
	\mathcal{H}_{\rm spin} = \sum_{i, j} J_{ij} \, \vec{S}_i \cdot \vec{S}_j
\end{equation}
defined on a lattice with arbitrary interactions $J_{ij}$ between $O(3)$ spins on sites $i$ and $j$.
We can rewrite this Hamiltonian as a sum over real-space coordinates 
\begin{equation}
	\mathcal{H}_{\rm spin} = \frac{1}{2} \sum_{\vec{r}} \sum_{A,B} \sum_{\vec{\mu}}  
						J_{AB}({\vec{\mu}}) \,\,  \vec{S}^{A}(\vec{r}\,) \cdot \vec{S}^{B}(\vec{r}+\vec{\mu} \,) \,,
	\label{eq:realspace-spin}
\end{equation}
where $\vec{r}$ runs over all unit cells, the indices $A$ and $B$ indicate the sites within the unit cell (for a non-Bravais lattice),
and the vectors $\vec{\mu}$ run over all connections between coupled sites.
To obtain the energy minima of this Hamiltonian one first performs a component-wise Fourier transformation to momentum space
yielding for each component
\begin{equation}
	\mathcal{H}_{\rm spin} = \sum_{\vec{k}} \sum_{A,B} S^{A}_{\vec{k}} \;\mathbf{M}_{A,B}(\vec{k})\; S^{B}_{-\vec{k}} \,,
	\label{eq:kspace-spin}
\end{equation}
where $\mathbf{M}_{A,B}(\vec{k})$ is the Fourier-transformed interaction matrix 
\begin{equation}
	\mathbf{M}_{A,B}(\vec{k}) = \frac{1}{2} \sum_{\vec{\mu}}  J_{AB}({\vec{\mu}}) \; e^{-i\vec{k}\vec{\mu}}
	\label{eq:heisenberg-matrix}
\end{equation}
of dimensionality $n \times n$ for a lattice with $n$ sites per unit cell. 
This Fourier-transformed interaction matrix can be readily diagonalized allowing to identify the momenta $\vec{k}$ with minimal energy eigenvalues. This manifold of minimal $\vec{k}$-vectors can be captured via
\begin{equation}
	\det\left(\mathbf{M}(\vec{k}) - E \cdot \mathbf{1} \right) = 0 \,,
	\label{eq:heisenberg-condition-eigenvalues}
\end{equation}
where $E$ is the ground-state energy.
The corresponding eigenvectors allow to reconstruct the real-space spin configurations via an inverse Fourier transformation,
which for eigenvectors obeying the original hard spin constraint generically yields a {\em coplanar} spin spiral of the form
\begin{equation}
\vec{S}^{A}(\vec{r}) = \operatorname{Re} \left( \left(\, \vec{u} + i\vec{v} \, \right)  S^{A}_{\vec{k}} e^{i\vec{k}\vec{r}} \right)  \,,
\end{equation}
where the vectors $\vec{u}$ and $\vec{v}$ span an arbitrary plane in the $O(3)$ spin space.

\subsubsection*{Tight-binding calculation for free-fermions}

The diagonalization of a real-space Hamiltonian  
via a Fourier transformation to momentum space is, of course, a rather
well known procedure frequently employed in the study of lattice systems.
Probably the most elementary example is its application to the solution of  
free-fermion Hamiltonians, which in real space take the tight-binding form
\begin{equation}
	\mathcal{H}_{\rm fermion} = \sum_{ i,j } t_{ij}  \, c_i^\dagger c^{\phantom\dagger}_j + {\rm h.c.} \,,
\end{equation}
where $c_i^\dagger$ and $c_j$ are fermionic creation and annihilation operators (acting at sites $i$ and $j$, respectively) 
and the $t_{ij}$ indicate the hopping strength.
Applying a Fourier transformation allows to write this Hamiltonian in momentum space as
\begin{equation}
	\mathcal{H}_{\rm fermion} = 
		\sum_{\vec{k}} \sum_{A,B} c^{\dagger}_{A,\vec{k}} \;\mathbf{H}_{A,B}(\vec{k})\; c^{\phantom\dagger}_{B,\vec{k}} \,,
\end{equation}
where the Fourier-transformed hopping matrix takes the form
\begin{equation}
	\mathbf{H}_{A,B}(\vec{k}) = \sum_{\vec{\delta}}  t_{AB}({\vec{\delta}}\,) \; e^{-i\vec{k}\vec{\delta}} \,,
	\label{eq:hopping-matrix}
\end{equation}
which is again an $n \times n$ matrix for a lattice with $n$ sites in the unit cell.
The Fermi surface (at $E_F = 0$ for the half-filled model at hand) can then be readily identified with the manifold of  $\vec{k}$-vectors satisfying
\begin{equation}
	\det\left(\mathbf{H}(\vec{k})\right) = 0 \,.
	\label{eq:fermisurface-condition-eigenvalues}
\end{equation}
Note that for the fermionic system there is no additional constraint on the eigenvectors -- each eigenstate at the Fermi energy
constitutes a valid solution. 

\subsection{Matrix correspondence}

As presented above, there is a close analogy between each step in the  Luttinger-Tisza (LT) approach 
and the tight-binding calculation. After Fourier transformation of the original real-space Hamiltonians 
both yield $n \times n$ matrices in momentum space that have matching forms, 
see Eqs. \eqref{eq:heisenberg-matrix} and \eqref{eq:hopping-matrix}. Diagonalizing these matrices yields
in both cases a momentum-resolved energy spectrum with $n$ bands. 
But while in the LT approach one is interested in the global energy {\em minimum} and the manifold of spin spiral
states defined by \eqref{eq:heisenberg-condition-eigenvalues}, the Fermi surface physics of the fermion
model plays out in the {\em middle} of the energy spectrum \eqref{eq:fermisurface-condition-eigenvalues}.
As such one cannot expect that for any given lattice the spin spiral manifold and the Fermi surface coincide
when simply identifying the interaction parameters $J_{ij}$ and $t_{ij}$ of the corresponding spin and fermion
lattice models.

To establish an identification of a spin spiral surface with a Fermi surface, one thus has to go an extra step. 
The crucial idea is to {\em square} the fermion interaction matrix
\begin{equation}
	\mathbf{H}(\vec{k}), \;  \varepsilon_j \quad \mapsto \quad \mathbf{H}(\vec{k})^2, \; \varepsilon_j^2 \;\;,
\end{equation}
which squares all its eigenvalues $\varepsilon_j$ and thereby moves its zero-energy eigenvalues to the bottom 
of the spectrum. In addition, one can shift these newly constructed minimal eigenvalues of the squared fermion matrix
by an arbitrary constant $E_0$ 
\begin{equation}
	\mathbf{H}(\vec{k})^2, \;  \varepsilon_j^2 \quad \mapsto \quad \mathbf{H}(\vec{k})^2 - E_0 \cdot \mathbf{1}, \;  \varepsilon_j^2 - E_0 \,.
\end{equation}
If we now consider a spin system whose interaction matrix precisely matches this latter form
\[
	{\bf M}(\vec{k}) \equiv {\bf H}(\vec{k})^2 - E_0 \cdot {\bf 1}  \,,
\]
then all minimal eigenvalues of ${\bf M}(\vec{k})$ correspond exactly to the zero eigenvalues of ${\bf H}(\vec{k})$. As such, the spin spiral surface defined by \eqref{eq:heisenberg-condition-eigenvalues} matches precisely the Fermi surface of \eqref{eq:fermisurface-condition-eigenvalues} and the ground-state energy of the spin system equals $E_0$.

This correspondence on the level of the interaction matrices is of rather universal character as it lays a general connection between a minimization problem (finding the ground state of a classical spin system) and the well known Fermi surface physics of free fermions. 
However, on this general level it does not readily imply a precise recipe for how to identify the spin and fermion {\em lattice models} underlying the  interaction matrices on the left and right hand side of this matrix correspondence.


\subsection{Lattice construction}
\label{sec:LatticeConstruction}

One way to realize the spin fermion correspondence \eqref{eq:MatrixCorrespondence} is to explicitly compose matching pairs 
of spin and fermion lattice models whose interaction matrices by construction satisfy the matrix correspondence. In the following, 
we will discuss an explicit lattice construction that works both ways -- either by starting from a given fermion lattice model and constructing the corresponding spin model or, vice versa, starting from a given spin model and constructing the corresponding fermion model. 

\subsubsection*{Fermions to spins}

To set the stage, let us start with a free fermion model on a given lattice and ask how we can explicitly construct a classical
spin model so that their respective interaction matrices will satisfy the matrix correspondence \eqref{eq:MatrixCorrespondence}.
The key observation of that correspondence is that one has to square the fermion matrix to match it to a spin model. This squaring
has a direct interpretation in terms of the lattices underlying the pair of corresponding spin and fermion models -- the lattice underlying the classical spin model is simply given by the next-nearest neighbor lattice of the fermion model as we will argue in the following.

\begin{figure}
	\centering
	\includegraphics[width=\columnwidth]{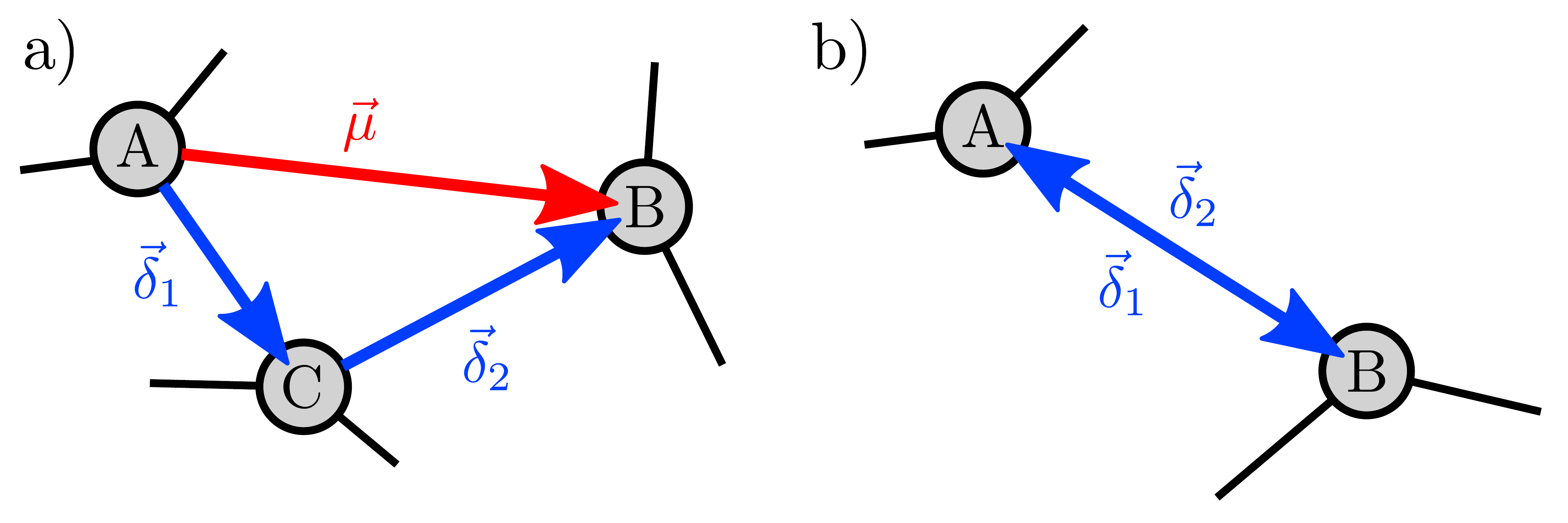}
	\caption{{\bf Spin lattice construction}. 
				Mapping the lattice of hopping fermions on a lattice of interacting spins by squaring the hopping 				
				matrix yields chained connections These can be interpreted as a new connection. 
				Diagonal elements are handled with care.
	}
	\label{fig:mapping-lattices-fermion-to-spin-ACA}
\end{figure}

To be specific, consider an element of the squared fermionic hopping matrix \eqref{eq:hopping-matrix} given by
\begin{equation}
	(\mathbf{H}^2)_{A,B}(\vec{k}) = 
	\sum_{C} \sum_{\vec{\delta}_1,\vec{\delta}_2}  t_{AC}({\vec{\delta}_1}) \, t_{CB}({\vec{\delta}_2}) \; 
						e^{-i\vec{k}(\vec{\delta}_1+\vec{\delta}_2)} \,,
	\label{eq:hopping-matrix-squared}
\end{equation}
which describes a hopping process between sites $A$ and $B$ through all possible intermediate sites $C$ as illustrated in Fig.~\ref{fig:mapping-lattices-fermion-to-spin-ACA}. In the language of the fermion model, this process describes a next-nearest neighbor hopping. Our goal here, however, is to interpret this matrix element as an element of the to-be-constructed spin interaction matrix \eqref{eq:heisenberg-matrix}. This can be readily accomplished by identifying
\begin{equation}
	J_{AB}({\vec{\mu}})  = 2 \sum_{C} t_{AC}({\vec{\delta}_1}) \, t_{CB}({\vec{\delta}_2}) \,,
	\label{eq:CorrespondenceCouplings}
\end{equation}
where $ \vec{\mu}  =  \vec{\delta}_1 + \vec{\delta}_2 	$. 
Note that one has to carefully distinguish the cases of $\vec{\mu} \neq 0$ and $\vec{\mu} = 0$.
For $\vec{\mu} \neq 0$ the fermionic hopping process always leads to a next nearest neighbor site
and can therefore always be identified as a contribution to the spin interaction matrix $\mathbf{M}(\vec{k})$.
Diagonal/off-diagonal matrix elements naturally arise if the labels of the connected next nearest neighbor sites
are equal/distinct (with the former case requiring the two sites to reside in different unit cells).
In contrast, for $\vec{\mu} = 0$ the fermion hopping describes a process returning to the original site as illustrated in Fig.~\ref{fig:mapping-lattices-fermion-to-spin-ACA}b). This generically leads to a $\vec{k}$-independent diagonal matrix
with elements
\begin{equation}
	\mathbf{E}^*_{AA} = \sum_{C, \vec{\delta}} \left| t_{AC}({\vec{\delta}}\,) \right|^2 \,,
	\label{eq:GoundStateEnergy}
\end{equation}
that describe all hopping processes from a site to its neighbors and back.

The squared fermionic interaction matrix thus naturally decomposes into the spin interaction matrix and an additional $\vec{k}$-independent diagonal matrix
\begin{equation}
	\mathbf{H}^2(\vec{k}) = \mathbf{M}(\vec{k}) + \mathbf{E}^* \,.
	\label{eq:square-relation-fermions-squared}
\end{equation}
By construction the spin interaction matrix $\mathbf{M}(\vec{k})$ is restricted to the lattice spanned by the next nearest neighbor bonds of the original fermion lattice. For a bipartite lattice this leads to a {\em decomposition} of the lattice into its two sublattices
and the newly constructed spin model is restricted to these individual sublattices. An elementary example is the fermionic honeycomb model, which decomposes into two triangular sublattices which underly the newly constructed spin model. 
Turning to the $\vec{k}$-independent diagonal matrix $\mathbf{E}^*$ we observe that this matrix typically has identical diagonal elements. For a large number of lattices this traces back to the fact that lattice symmetries require all sites to have the same local connectivity. For these lattices, the diagonal matrix $\mathbf{E}^*$ can be expressed as $\mathbf{E}^* = E_0 \cdot {\bf 1}$, where by construction $-E_0$ must be equal to the lowest eigenvalue of the spin interaction matrix $\mathbf{M}(\vec{k})$, since they sum to zero -- the lowest eigenvalue of the squared fermion matrix in Eq.~\eqref{eq:square-relation-fermions-squared}. This implies that $-E_0$ is the ground-state energy of the spin model. In our example sections, we will return to this point and discuss several instances of this correspondence including cases where multiple distinct entries in the diagonal matrix $\mathbf{E}^*$ occur.


\subsubsection*{Spins to fermions}
\label{sec:Construction}

We now turn to the inverse direction and ask how one can explicitly construct a fermion model from a given spin model. 
In the language of the matrix correspondence \eqref{eq:MatrixCorrespondence} this corresponds to ``taking the square root"
of the spin interaction matrix. While this might be a formidable task, we note that in the fermion to spin model construction,
the squaring of the fermion matrix translated into an explicit lattice construction. It is precisely this lattice
construction that allows for a relatively straight-forward inversion and thereby allows us to lay out an explicit procedure
how to construct a fermion model as the ``square root" of a given spin model.

\begin{figure}[t]
	\centering
	\includegraphics[width=\columnwidth]{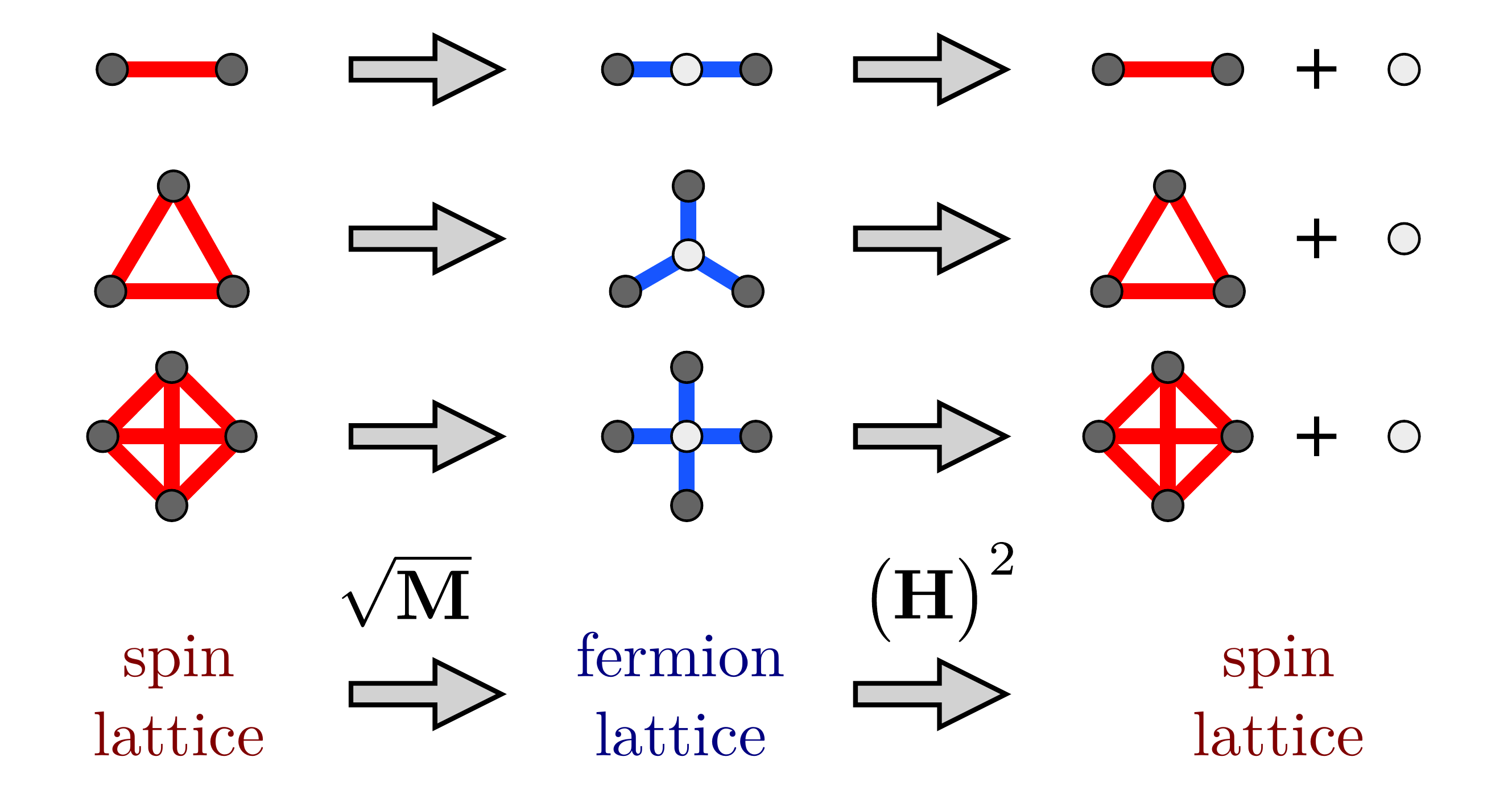}
	\caption{{\bf Fermion lattice construction}. Mapping plaquettes of the fermion lattice (middle) to the resulting parts in the spin lattice (right). One can easily see that in a bipartite lattice all sites of one type form one individual sublattice. Therefore, this procedure is taken as a controlled way to map a spin lattice to a fermion lattice}
	\label{fig:mapping-spin-to-fermion}
\end{figure}

In the fermion to spin model construction (described in the previous subsection) we have seen that for bipartite fermion lattices
the squaring of the hopping matrix ${\bf H}$ has led to a decomposition of the original lattice into its sublattices. In particular, 
as illustrated in Fig.~\ref{fig:mapping-spin-to-fermion}, this decoupling transforms elementary $z$-coordinated sites of the fermion lattice into fully connected plaquettes of $z$ sites in the corresponding spin lattice. 
Thus, in an inverse construction going from a spin to a fermion model one has to invert precisely this step by replacing all fully connected plaquettes of $z$ sites by newly added $z$-coordinated sites. Noting that fully connected plaquettes of $z$ sites naturally contain all fully connected plaquettes of smaller sizes (e.g. the checkerboard plaquette in the bottom row of Fig.~\ref{fig:mapping-spin-to-fermion} contains four triangular plaquettes), one has to replace the  fully connected plaquettes of any given spin lattice by decreasing size. 
Further note that the insertion of additional sites  
upon replacing  fully connected plaquettes in the spin lattice is precisely what gives rise to the bipartite structure of the fermion lattice. Note that these added sites constitute the second sublattice that does not necessarily have be to identical to the original spin lattice. 
This readily implies a deeper connection also between these two spin models -- a point to which we will return below.

This geometric construction of the fermion model is a fully consistent inversion of the fermion to spin model construction, 
i.e. by subsequently applying both lattice constructions one indeed returns to the orginal spin or fermion model.
This is illustrated for the elementary lattice motives in Fig.~\ref{fig:mapping-spin-to-fermion} and fleshed out for a large number of lattices in the example sections.


\subsection{Symmetry class classification}
\label{sec:SymmetryClass}

One instructive way to think about the spin fermion correspondence laid out in this manuscript, is to classify the types of
free fermion models which can arise in this correspondence in terms of the 10-fold way symmetry class classification \cite{Altland1997}. 
To do so, we note that the mapping of a spin model to a fermion model in our correspondence leads to a free spinless
fermion model on a bipartite lattice. The fermionic hopping matrix will take the general block form 
\begin{equation}
	\mathbf{H} = \begin{pmatrix}
				0 & \mathbf{Q} \\
				\mathbf{Q}^\dagger & 0
			      \end{pmatrix} \,,
	\label{eq:FermionMatrixBlocks}
\end{equation}
where the two off-diagonal subblocks correspond to the hopping matrices between the two sublattices. 
In general, these blocks are matrices of dimensionality ${n\times m}$, where $n$ and $m$
indicate the number of sites in the unit cells of the two sublattices (and which do not necessarily have to be identical).
With this matrix form the fermion model naturally falls into one of the chiral symmetry classes. Upon further inspection,
one finds that since the Hamiltonian exhibits time-reversal symmetry, particle-hole symmetry (at half-filling), and
sublattice symmetry. As such the fermion system generically resides in symmetry class BDI of the 10-fold way classification \cite{Altland1997}.

\subsubsection*{Further symmetry considerations}

In passing we note that squaring a free fermion matrix of the general form \eqref{eq:FermionMatrixBlocks} yields
a  block diagonal matrix
\begin{equation}
	\mathbf{H}^2 = \begin{pmatrix}
					\mathbf{Q}\mathbf{Q}^\dagger & 0 \\
					0 & \mathbf{Q}^\dagger \mathbf{Q}
				\end{pmatrix}
\end{equation}
that via our matrix correspondence is directly related to the spin interaction matrix ${\bf M}$. 
The two blocks $\mathbf{Q}\mathbf{Q}^\dagger$ and $\mathbf{Q}^\dagger \mathbf{Q}$ capture 
the spin interactions on the two sublattices of the fermion lattice, respectively.
Note that for the case $m\neq n$, i.e. when the two sublattices have unit cells of different sizes, 
then the spin models with the larger number of sites per unit cell will have $|m-n|$ flat bands at its ground-state energy.
Analogously, the respective fermion model will have a corresponding number of flat bands at the Fermi energy.
This can be seen from the fact that if $m\neq n$, then the matrix rank of $\mathbf{Q}$ is the minimum of $m$ and $n$, 
and the higher-dimensional block of $\mathbf{Q}\mathbf{Q}^\dagger$ and $\mathbf{Q}^\dagger \mathbf{Q}$ will have
a kernel of dimension $|m-n|$. Like before, these zero-energy eigenvalues (of the fermion hopping matrix) will be shifted 
to match the ground-state energy of the respective spin model via our matrix correspondence.


\section{Case studies}
\label{sec:CaseStudies}

We now turn to a number of case studies illustrating the spin fermion correspondence.
Our guiding motif in choosing these examples is to start from a frustrated Heisenberg model
on some magnetic lattice and to construct the corresponding fermion lattice model. 
We start with the simplest scenario -- frustrated magnets defined on Bravais lattices such as the triangular
lattice or the fcc lattice. We then turn to non-Bravais lattices, for which additional care in checking the
validity of the Luttinger-Tisza approach is needed. Specifically, we consider the Heisenberg models on the
kagome and pyrochlore lattices.
As an additional variation we consider $J_1-J_2$ Heisenberg models with both nearest and next-nearest neighbor 
spin exchange on a variety of lattices.

Going through these examples we will present (multiple) instances of frustrated magnets where the ground-state
spin spiral manifold is captured by points, lines, surfaces, or entire Brillouin zone volumes. 
A summary is given in Table \ref{tab:overview} of the Introduction.


\subsection{Magnetic Bravais lattices}

As a first set of examples we consider frustrated Heisenberg antiferromagnets defined on Bravais lattices.
For these particularly simple lattices, the Luttinger-Tisza approach is known to be exact \cite{Lyons1960},
i.e. all minimal-energy states must satisfy the hard spin constraint and therefore capture a valid ground state 
of the magnetic system.


\subsubsection{Triangular lattice} 

\begin{figure}[b]
	\centering
	\includegraphics[width=\linewidth]{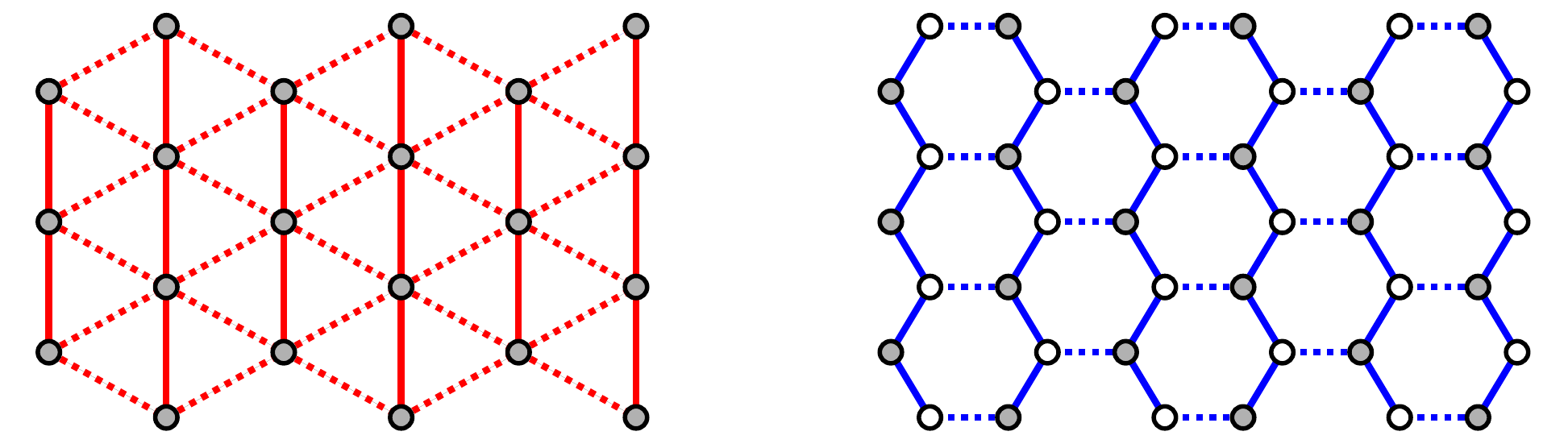}	
	\caption{The {\bf triangular} lattice with spatially anisotropic couplings indicated by the bold and dashed lines.
			The corresponding fermion model is defined on the honeycomb lattice with similarly spatially anisotropic couplings.}
	\label{fig:honeycomb-lattice-fermions-aniso}
\end{figure}

A most illustrative example to start with is the triangular lattice Heisenberg antiferromagnet. 
Its corresponding fermion model is defined on the honeycomb lattice, which follows directly 
from the lattice construction discussed in Section \ref{sec:LatticeConstruction}. This can be 
easily seen by inspecting the two lattices as illustrated in Fig.~\ref{fig:honeycomb-lattice-fermions-aniso}.
Starting from the fermionic honeycomb lattice, one immediately finds that the sublattices spanned
by next-nearest neighbor bonds are indeed triangular lattices. Going in the other direction works 
in a similarly straight-forward way by replacing all up-pointing triangles in the triangular lattice 
by tricoordinated sites,
which immediately yields the honeycomb lattice. Note that only one type of triangles needs to
be replaced in this reverse lattice construction as this will already suffice to replace all bonds 
in the original triangular lattice.

In the spirit of the spin fermion correspondence we can now proceed to discuss the ground state of the triangular
lattice antiferromagnet by identifying it with the well known Dirac physics of free fermions on the honeycomb lattice.
To do this, let us quickly recap the elementary free-fermion calculation on the honeycomb lattice in the language
of the previous Section. Considering isotropic hopping along all nearest neighbor bonds the hopping matrix for the
honeycomb lattice with its two-site unit cell takes the well-known form
\begin{equation}
	\mathbf{H}(\vec{k}) = \begin{pmatrix} 0 & f(\vec{k}) \\ f^*(\vec{k}) & 0 \end{pmatrix} \,,
	\label{eq:Dirac}
\end{equation}
where the off-diagonal matrix elements are given by 
$	f(\vec{k}) = t \left( e^{-i \vec{k} \vec{\delta}_1} + e^{-i \vec{k} \vec{\delta}_2} + e^{-i \vec{k} \vec{\delta}_3} \right) $ 
with nearest-neighbor bonds along $\vec{\delta}_1 = (1/\sqrt{3}, 0)$ and $\vec{\delta}_{2,3} = (-1/2\sqrt{3},\pm 1/2)$.
Diagonalizing the hopping matrix ~\eqref{eq:Dirac} gives the two-band energy spectrum
$	\varepsilon_{\pm}(\vec{k}) = \pm \sqrt{f^*(\vec{k})f(\vec{k})} $
with its two characteristic Dirac cones located at 
\begin{equation}
	\vec{k}_{1,2} = \left(\frac{2\pi}{\sqrt{3}}, \pm \frac{2\pi}{3}\right) \,.
	\label{eq:kpoints-honeycomb}
\end{equation}
The nodal structure of the honeycomb fermion model is thus constituted by precisely these two points.


We can now turn to the triangular lattice antiferromagnet. Quickly going through the Luttinger-Tisza approach
we can write down the interaction ``matrix" ${\bf M}(\vec{k})$, which for the triangular lattice with its single site 
in the unit cell reduces to a real function. It is explicitly given by
\begin{align*}
	{\rm M}(\vec{k}) &= J \big( \cos( \vec{k}(\vec{a}_1 + \vec{a}_2)/3)  
	+ \cos(\vec{k}(-2\vec{a}_1 + \vec{a}_2)/3)  \\ 
	& \quad \quad \quad + \cos(\vec{k}(\vec{a}_1 -2 \vec{a}_2)/3) \big) \,.
\end{align*}
The minima of this function, $E_0 = -3/2~J$, are located at the two $\vec{k}$-points $\left({2\pi}/{\sqrt{3}}, \pm {2\pi}/{3}\right)$.
These two minima describe two spin spiral states -- the two possible orientations of the well-known 120$^\circ$ order of the triangular lattice antiferromagnet.

With these two elementary calculations in place, we can now see how the spin fermion correspondence plays out in this example.
The 120$^\circ$ order of the triangular lattice antiferromagnet is captured by precisely the two $\vec{k}$-points that indicate the
location of the Dirac cones of the fermionic honeycomb model. That is, the spin spiral manifold of the triangular lattice model is
precisely captured by the nodal manifold of the fermionic honeycomb model.

To also establish this connection also on the level of the original matrix correspondence, we note that $\mathbf{H}(\vec{k})^2$
is a $2 \times 2$ matrix of the block-diagonal form
\begin{eqnarray}
	\mathbf{H}(\vec{k})^2 & = & \begin{pmatrix}  |f(\vec{k})|^2 & 0\\ 0 &  |f(\vec{k})|^2 \end{pmatrix} \nonumber \\
					& \equiv & \begin{pmatrix} {\rm M}(\vec{k})-E_0 & 0 \\ 0 & {\rm M}(\vec{k})-E_0 \end{pmatrix} \,,
	\label{eq:TriangularMatrixCorrespondence}
\end{eqnarray}
where each diagonal block corresponds to the description of one of the two triangular sublattices.
Note that one could have derived the numerical value of the ground-state energy $E_0$ from this correspondence via Eqs.~\eqref{eq:CorrespondenceCouplings} and \eqref{eq:GoundStateEnergy} {\em without} minimizing ${\rm M}(\vec{k})$.


\subsubsection*{Anisotropic exchange}

\begin{figure}[t]
	\centering
	\includegraphics[width=0.94\linewidth]{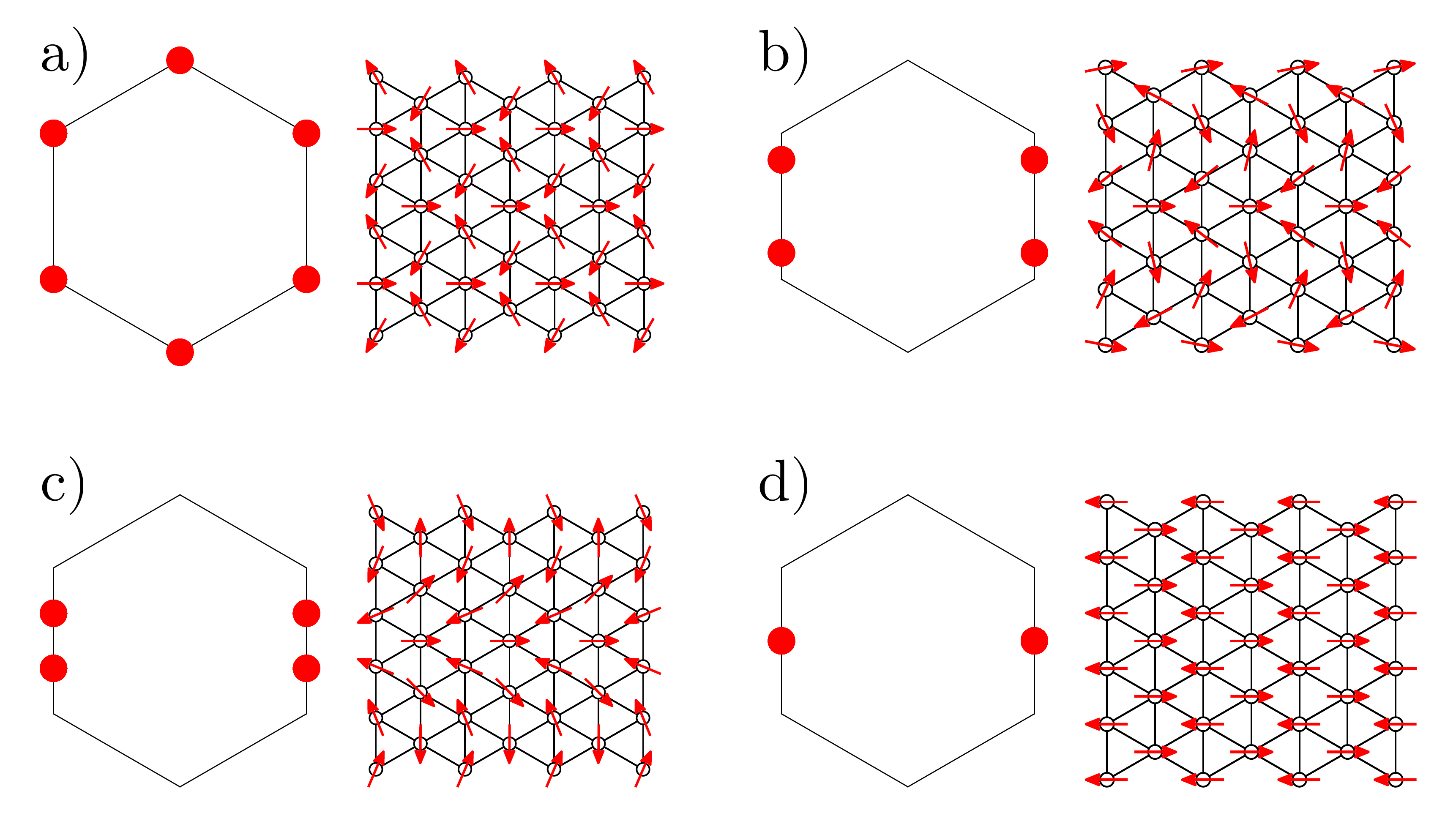}
	\caption{Ground state structure of the spin model on the triangular lattice for {\bf anisotropic coupling}. Varying the hopping amplitude $t_{-}$ from $t_{-} = 1/3$ to $t_{-} = 1/2$ merges the 120$^\circ$ order points along the edges of the Brillouin zone. For $t_{-} > 1/2$ the fermion system develops a gap so that the Fermi surface vanishes but the spin system stays in an antiferromagnetic alingment in the groundstate. Besides the appearing k-points, a visualization of the groundstate spiral on the lattice is given.}
	\label{fig:GS-triangular-aniso}
\end{figure}

For the fermionic honeycomb model it is well known that one can move the location of the Dirac points
by introducing a spatially anisotropic hopping. For instance, strengthening the hopping along the horizontal bonds
in the honeycomb lattice (indicated by the dashed lines in Fig.~\ref{fig:honeycomb-lattice-fermions-aniso})
moves the two Dirac cones towards each other along the line connecting the $K$ and $K^\prime$ points in
the Brillouin zone. Via our spin fermion correspondence this immediately implies that an anisotropic spin exchange
in the triangular lattice antiferromagnet has the exact same effect -- the $\vec{k}$-vectors describing the two
coplanar spiral ground state configurations move towards one another as illustrated in Fig.~\ref{fig:GS-triangular-aniso}.

To be more specific, let us denote with $t_{-}$ and $t_{\rangle}$ the hopping along the horizontal and vertical zigzag bonds
in the fermionic honeycomb lattice, respectively. The corresponding triangular spin model then exhibits an anisotropic spin 
exchange with coupling constants $J_{|}$ and $J_\times$ along the vertical and diagonal coupling directions, respectively.
The spin and fermion couplings are related to one another via Eq.~\eqref{eq:CorrespondenceCouplings}
\[
	J_{\times} = 2 \, t_{-} t_{\rangle} \quad {\rm and} \quad J_{|} = 2 \, t_{\rangle}^2 \,\,.
\]
With this relation of the couplings in place, the general form of the matrix correspondence \eqref{eq:TriangularMatrixCorrespondence} holds for all values of the couplings.
Notably, this is in particular the case for $t_{-} > 2 t_{\rangle}$ where the fermionic system exhibits a {\em gap} in the
excitation spectrum. For the corresponding spin couplings the magnetic ordering remains fixed in the N\'eel state,
captured by $\vec{k} = ({2\pi}/{\sqrt{3}},0)$ -- the momentum which corresponds precisely to the location of the gap
in the fermionic band structure. By virtue of the matrix correspondence \eqref{eq:TriangularMatrixCorrespondence}
one can directly calculate the size of the gap in the fermionic band structure as $\Delta E = 2 \sqrt{E^* - E_0} = 4 (t_{-} - 1/2) $.


\subsubsection{FCC lattice}

\begin{figure}[t]
	\centering
	\includegraphics[width=\columnwidth]{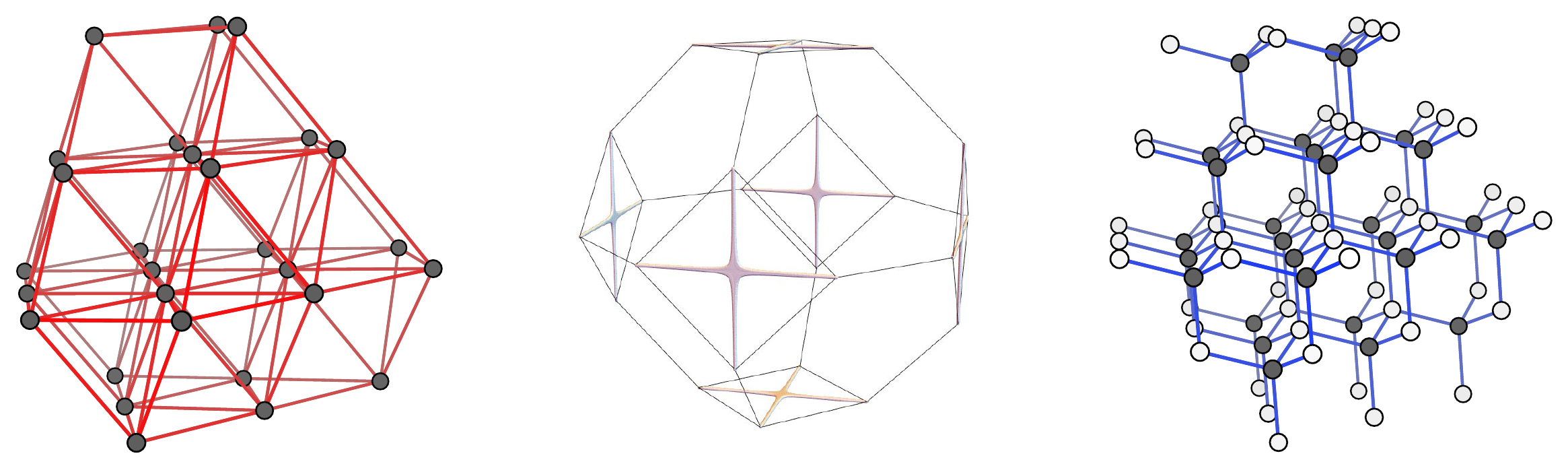}
	\caption{The geometrically frustrated Heisenberg antiferromagnet on the {\bf FCC lattice} (left) is 
			related via the spin fermion correspondence to free fermions on the diamond lattice (right).
			Their common spiral/Fermi surface is shown in the middle panel.}
	\label{fig:DiamondLattice}
\end{figure}

As a second example of a magnetic Bravais lattice we consider the Heisenberg antiferromagnet on the three-dimensional 
face-centered cubic (FCC) lattice. Its corresponding fermion model lives on the diamond lattice, which can be easily seen by remembering that the diamond lattice consists of two FCC sublattices. With the illustration of Fig.~\ref{fig:DiamondLattice} at hand,
one can also see the inverse lattice correspondence -- if one replaces every up-pointing tetrahedron in the FCC lattice by a four-coordinated site one instantly arrives at the diamond lattice.

Turning to the spin spiral / nodal manifold of the respective models, we find that these are described by two crossing {\em lines} 
confined to the square-shaped faces of the Brilouin zone as illustrated in the middle panel of Fig.~\ref{fig:DiamondLattice}.

\subsection{Magnetic non-Bravais lattices}
\label{sec:ExamplesNonBravais}

In considering examples for our spin fermion correspondence a natural next step is to consider Heisenberg antiferromagnets on non-Bravais lattices. The latter include some quintessential frustrated magnets such as the kagome antiferromagnet in two spatial dimensions or the pyrchlore anti\-ferromagnet in three spatial dimensions. Both stand out as they are known to exhibit an {\em extensive} ground-state degeneracy. For both lattices, the source of this extensive degeneracy can be tracked back to the lattice geometry of corner-sharing triangle or tetrahedra, respectively, which gives rise to a {\em local} constraint on the total spin of each triangle/tetrahedron \cite{Moessner1998}.
We will consider precisely these two examples in the following and explicitly construct their corresponding fermion models. 


\begin{figure}
	\includegraphics[width=\columnwidth]{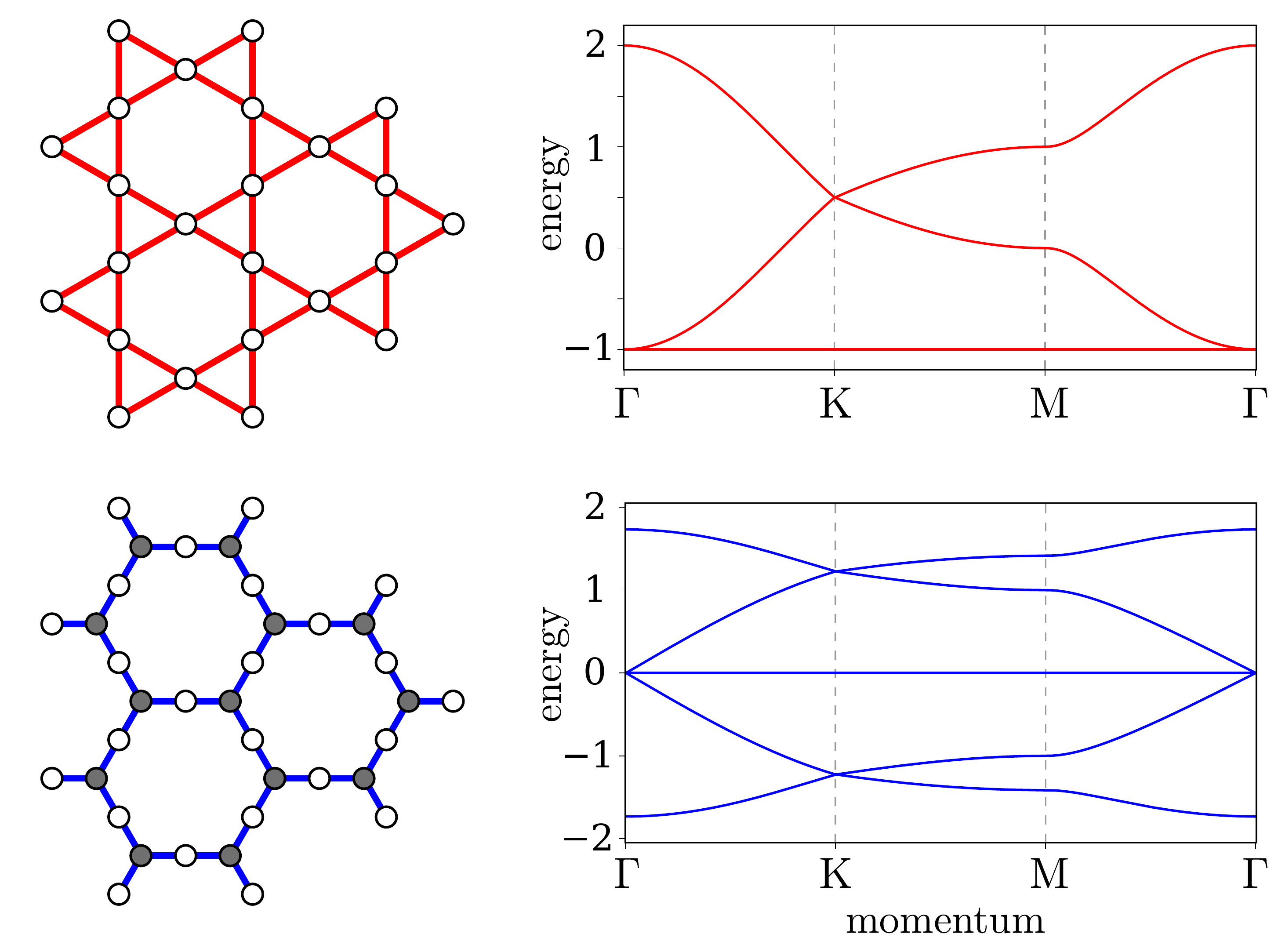}
	\caption{Spectrum and flat bands of the spin model on the {\bf kagome lattice} 
			as well as of the corresponding fermion model on an extended honeycomb lattice.
			The fermionic spectrum (bottom) is the positive/negative square root of the Luttinger-Tisza spectrum (top).
			Note the three-fold band crossing of a Dirac cone and a flat band at the $\Gamma$ point.}
	\label{fig:kagome}
\end{figure}

Let us first consider the kagome antiferromagnet. With a unit cell of three sites, one has to exercise care in performing a Luttinger-Tisza calculation as it is no longer guaranteed to result in valid ground states that obey the uniform spin length constraint. Without enforcing the hard spin constraint, one finds the Luttinger-Tisza spectrum \cite{Chalker1992,Huse1992,Harris1992} of Fig.~\ref{fig:kagome}.
The extensive ground-state degeneracy manifests itself in the occurrence of a {\em flat band} at the minimal energy in this spectrum.
The corresponding fermion model is readily constructed (via the lattice construction of Section \ref{sec:LatticeConstruction}) by replacing all elementary triangles in the kagome lattice by tricoordinated sites, which gives what we denote as the extended honeycomb lattice illustrated in Fig.~\ref{fig:kagome}. The fermionic spectrum contains five bands with a flat band residing precisely at the Fermi energy. 
The origin of the flat band can be traced back to the arguments presented in Section \ref{sec:SymmetryClass} and noting that the extended honeycomb lattice
decomposes into two {\em distinct} sublattices, a kagome and a honeycomb lattice, with differing number of sites in the unit cell. 
Note that the fermionic spectrum is indeed given by the positive/negative square root of the Luttinger-Tisza spectrum of the spin model.
Probably the most notable feature of this fermionic spectrum beyond the flat band is the Dirac cone crossing the flat band at the $\Gamma$ point -- an interesting example of a high degeneracy point (for a spinful fermion model there is a six-fold degeneracy).

\begin{figure}[t]
	\includegraphics[width=\columnwidth]{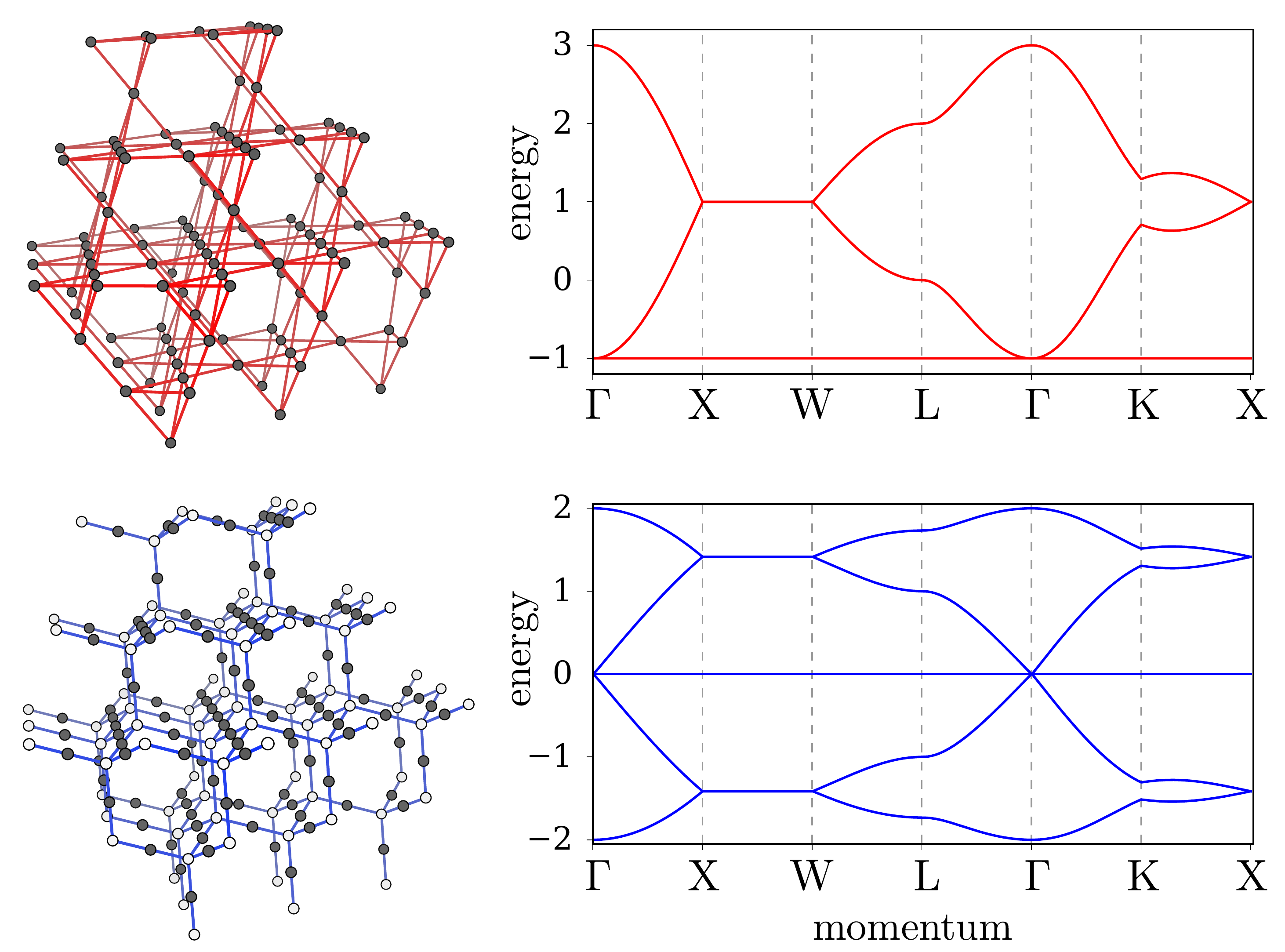}
	\caption{Spectrum and flat bands of the spin model on the {\bf pyrochlore lattice} 
			as well as of the corresponding fermion model on the diamond-x lattice.
			The fermionic spectrum (bottom) is the positive/negative square root of the Luttinger-Tisza spectrum (top).
			Note the four-fold band crossing of a Dirac cone and two flat bands at the $\Gamma$ point.}
	\label{fig:pyrochlore}
\end{figure}

A similar picture emerges for the pyrochlore lattice. With its four-site unit cell one obtains the Luttinger-Tisza spectrum \cite{Reimers1991} of Fig.~\ref{fig:pyrochlore}, which contains two degenerate flat bands at the minimal energy indicative of the extensive ground-state degeneracy. The corresponding fermion model is obtained by replacing every elementary tetrahedron in the pyrochlore lattice by a four-coordinated site resulting in the extended diamond lattice (composed of a pyrochlore and diamond sublattice) illustrated in Fig.~\ref{fig:pyrochlore}. Its spectrum is again given by the positive/negative square root of the Luttinger-Tisza spectrum. Similarly to the two-dimensional extended honeycomb model, probably its most notable feature 
 is an eight-fold degeneracy at the $\Gamma$ point (for a spinful fermion model) where a Dirac cones crosses the two degenerate flat bands. Such highly degenerate crossings (at high-symmetry points in the Brillouin zone) are of fundamental interest \cite{Wieder2016,Bradlyn2016} as they allow for elementary fermionic excitations beyond the standard (high-energy) classification of Dirac, Weyl, and Majorana fermions.

\subsection{$J_1$-$J_2$ Heisenberg models}
\label{sec:J1J2}

Another often-studied family of frustrated spin models are $J_1$-$J_2$ Heisenberg models where an antiferromagnetic next-nearest neighbor exchange $J_2$ destabilizes the conventional order of the nearest neighbor model. A particularly interesting member of this family is the $J_1$-$J_2$ Heisenberg model on the diamond lattice, which is one of the few frustrated magnets to 
exhibit a subextensive ground-state degeneracy captured by a spiral {\em surface} in momentum space \cite{Bergman2007}.
These spiral surfaces have been proposed to occur in certain $A$-site spinel compounds \cite{Bergman2007,Lee2008,Bernier2008,Chen2009,Savary2011,Chen2017}, with recent 
inelastic neutron scattering experiments on MnSc$_2$S$_4$ indeed reporting their unambiguous experimental observation  \cite{Gao2017}. In the context of this manuscript, the spiral surface of the $J_1$-$J_2$ diamond model bears, of course, 
the most striking resemblance to the Fermi surface of a metal. In the following, we will lay out how to construct a fermionic tight-binding model that exhibits precisely the same Fermi surface. We will also discuss a number of other $J_1$-$J_2$ Heisenberg models that exhibit spin spiral surfaces such as the $J_1$-$J_2$ honeycomb model or a modified $J_1$-$J_2$ model on the bcc lattice.

To be specific, we consider a model
\begin{equation}
	\mathcal{H} = J_1 \sum_{\langle i,j \rangle} \vec{S}_i \vec{S}_j + J_2 \sum_{\langle \langle i,j \rangle \rangle} \vec{S}_i \vec{S}_j 
	\,,
	\label{eq:heisenberg-J1J2}
\end{equation}
where $\langle i,j \rangle$ and $\langle \langle i,j \rangle \rangle$ denote nearest and next-nearest neighbors, respectively.
We will restrict ourselves to bipartite lattices in the following, i.e. lattices that generally decompose into two individual sublattices $A$ and $B$.
In terms of these sublattices, $J_1$ is a coupling (of arbitrary sign) {\em between} the two sublattices, while $J_2$ is an antiferromagnetic coupling  only {\em within} one of the two sublattices.  

\begin{figure}[t]
	\centering
	\includegraphics[width=\columnwidth]{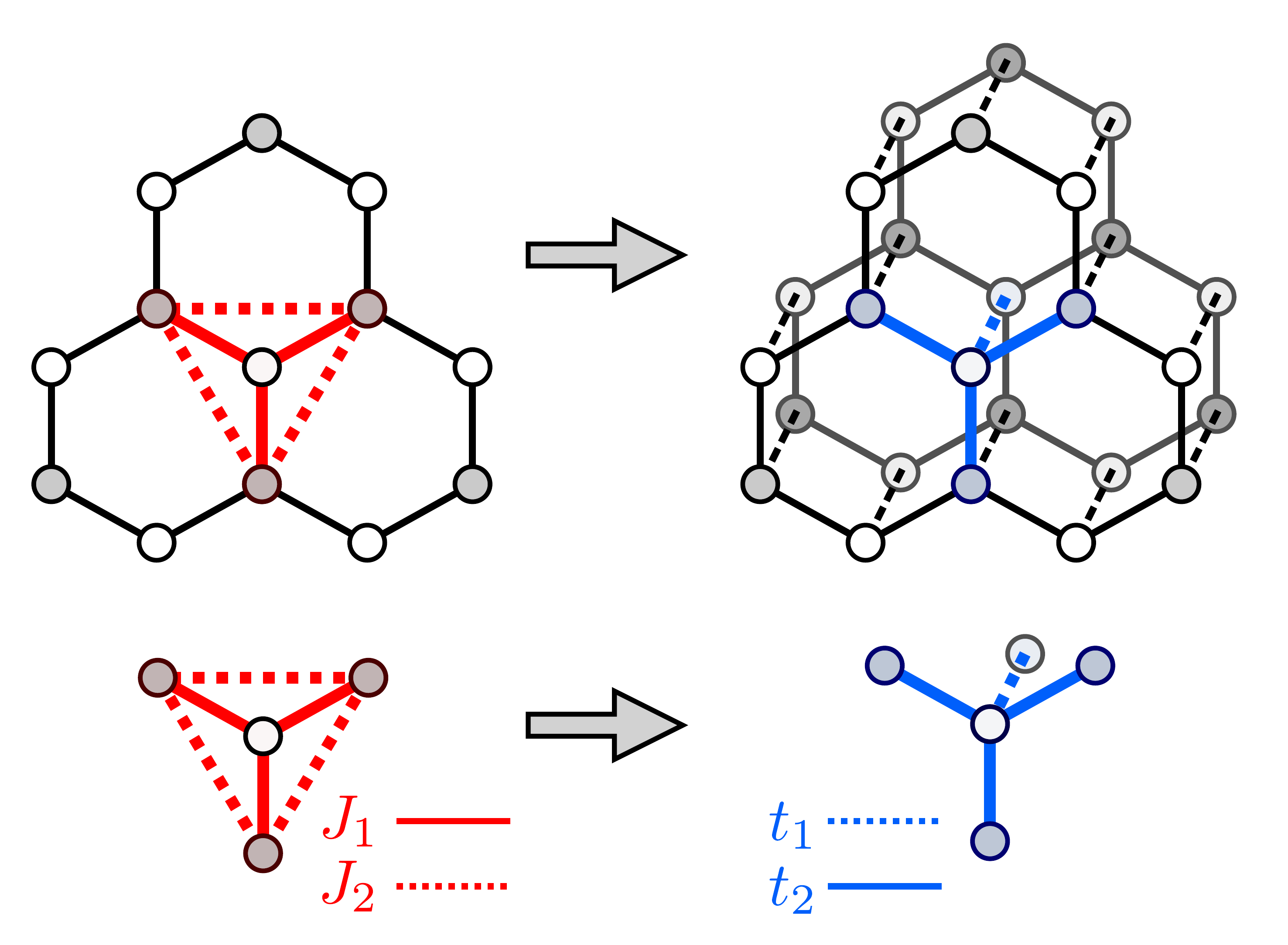}
	\caption{The {\bf spin fermion correspondence for $J_1$-$J_2$ Heisenberg models} illustrated for the honeycomb model.
		     In employing the lattice construction scheme of Fig.~\ref{fig:mapping-spin-to-fermion}, 
		     the elementary step is to replace each fully connected plaquette 
		     of nearest and next-nearest neighbor bonds around a given site (highlighted) as illustrated.
		     The end result is a fermionic hopping model on a bilayer of the original spin lattice.}
	\label{fig:J1J2-plaquette-mapping}
\end{figure}

To map such a $J_1$-$J_2$ spin model to a fermion model in terms of the spin-fermion correspondence, one can readily employ
the lattice construction algorithm of Section \ref{sec:LatticeConstruction}. Specifically, the elementary step is to again replace all fully connected plaquettes in the spin lattice by fermionic sites. The largest fully connected plaquette for a $J_1$-$J_2$ spin model on a $z$-coordinated lattice
can be easily identified. It is given by the $z$ sites around a given site connected via both nearest and next-nearest neighbor connections.
As an example, this is illustrated for the honeycomb $J_1$-$J_2$ model in Fig.~\ref{fig:J1J2-plaquette-mapping}. 
Note that this construction replaces all bonds of the original spin model in a single step of the lattice construction, if the next-nearest neighbor coupling $J_2$ is defined by {\em bond} distance (which in turn leads to the fully connected plaquettes described above). 
For systems where $J_2$ is defined more generally, e.g. by real-space distance or involving spatial anisotropies, the lattice construction might include multiple steps (which will not be discussed in the following).
Replacing all fully connected plaquettes as described above, the end result of the lattice construction generically is a fermionic {\em bilayer}  of two copies of the original spin lattice.

In terms of coupling constants, the spin fermion correspondence of Eq.~\eqref{eq:CorrespondenceCouplings} immediately gives
\begin{eqnarray}
t_1 &=& J_1 / \sqrt{8 J_2}\,, \nonumber \\
t_2 &=& \sqrt{J_2/2} \,,
\end{eqnarray}
where $t_1$ is the coupling {\em between} the two layers and $t_2$ is the coupling {\em within} a layer, see also Fig.~\ref{fig:J1J2-plaquette-mapping}. 
Note that in the limit of $J_1 = 0$, where the spin model decomposes into its two individual sublattices, the corresponding
 fermionic bilayer model likewise decomposes into two separate layers (with each layer corresponding to one of the sublattices 
in the spin model).

As an alternative to formulating the fermion model as a hopping Hamiltonian on a bilayer lattice, one can introduce a $Z_2$ spin variable (indicating the layer) so that one arrives at a spinful fermion Hamiltonian on the original spin lattice of the form
\begin{equation}
\mathcal{H} = t_2 \!\!  \sum_{\substack{\langle i,j \rangle \\ \sigma=\uparrow,\downarrow}} \! c_{j,\sigma}^\dagger c_{i,\sigma}^{\phantom\dagger} \; + \; t_1 \sum_{i} \left( c_{i,\uparrow}^\dagger c_{i,\downarrow}^{\phantom\dagger} + {\rm h.c.} \right) \,,
\label{eq:spinful-fermion-J1J2}
\end{equation}
where $t_2$ is the strength of a spin-conserving hopping on the lattice and $t_1$ now parametrizes an on-site spin-flip term.

Before we turn to a discussion of a number of example {$J_1$-$J_2$} spin models, let us note that it typically takes a {\em finite}
coupling strength $J_2$ to destabilize the N\'eel (or ferromagnetic) order favored by the nearest neighbor coupling $J_1$. 
The exact strength of the critical coupling can be derived directly from the spin-fermion correspondence by noting that the 
conventionally ordered magnetic phases correspond to a gapped fermion spectrum and the critical point coincides with the 
gap closing for increasing $J_2$. Starting from a N\'eel ordered state (for antiferromagnetic $J_1$) its ground-state energy reads
\[
	E_{\text{N\'eel}} = -\frac{1}{2} J_1 z + \frac{1}{2} J_2 z(z-1)
\]
as a function of $J_2$. On the other hand, the Fermi energy of the  corresponding fermion model maps according to Eq.~\eqref{eq:GoundStateEnergy} to
\[
	- E^* =  -  \frac{zJ_2}{2} - \frac{J_1^2}{8J_2} \,,
\]
which is lower than the above N\'eel energy. The gap $E_{\text{N\'eel}} + E^*$ closes precisely at $J_2/J_1 = 1/(2z)$,
which indeed indicates the transition to coplanar spin spiral ground states (with energy $- E^*$).  


\subsubsection*{Examples}

\begin{figure}[b]
	\centering
	\includegraphics[width=\columnwidth]{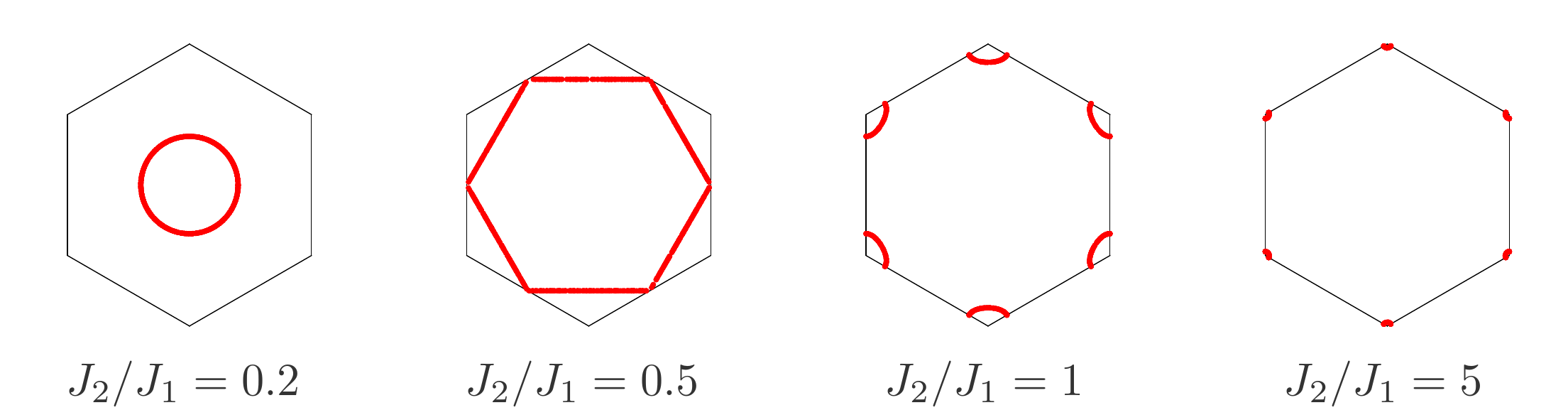}
	\caption{Spin spiral lines for the {\bf honeycomb lattice} 
			$J_1$-$J_2$ Heisenberg model for different ratios of the couplings.
			}
	\label{fig:HoneycombSpiralSurfaces}
\end{figure}

To illustrate the above spin fermion correspondence for $J_1$-$J_2$ Heisenberg models we proceed with a number of examples.
Starting in two spatial dimensions, the $J_1$-$J_2$ model on the honeycomb lattice might be of particular interest. As worked 
out in detail in Ref.~\cite{Mulder2010}, this model exhibits spin spiral ground states for $J_2/J_1 > 1/6$ captured by a {\em line}
in momentum space whose evolution with varying $J_2$ is illustrated in Fig.~\ref{fig:HoneycombSpiralSurfaces}. Note that the $J_2 \to \infty$ limit corresponds to the 120$^\circ$ order of the triangular lattice antiferromagnet discussed above.
In terms of the spin fermion correspondence, these spin spiral lines correspond to the Fermi surface of a spinful fermion model on the honeycomb lattice. Starting from the two Dirac cones in the $t_1=0$ ($J_2 \to \infty$) limit, an onsite spin-flip term leads to the formation of a full Fermi surface (i.e. lines in two spatial dimensions) by shifting the Dirac cones above and below the Fermi energy.

\begin{figure}[h]
	\centering
	\includegraphics[width=\columnwidth]{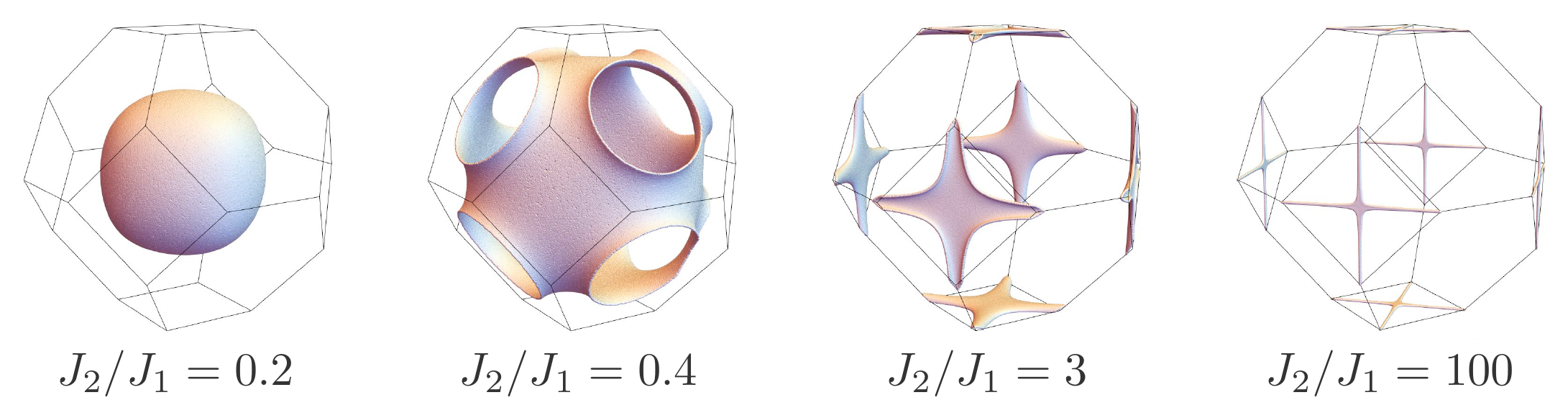}
	\caption{Spin spiral surfaces for the {\bf diamond lattice} 
			$J_1$-$J_2$ Heisenberg model for different ratios of the couplings.}
	\label{fig:DiamondSpiralSurfaces}
\end{figure}

In three spatial dimensions, we first turn to the $J_1$-$J_2$ diamond model of Ref.~\cite{Bergman2007}. For a next-nearest neighbor coupling $J_2/J_1 > 1/8$ this model exhibits a spin spiral surface that with increasing $J_2$ changes its topology as illustrated in Fig.~\ref{fig:DiamondSpiralSurfaces}. In the limit $J_2 \to \infty$ one recovers the physics of the fcc lattice whose ground-state 
spin spiral manifold is described by a set of crossing nodal lines \cite{Henley1987}. Via our spin fermion correspondence we can readily identify these spin spiral surfaces with the Fermi surfaces of fermions hopping on a bilayer diamond lattice, or analogously with a spinful fermion model \eqref{eq:spinful-fermion-J1J2} on the diamond lattice.

\begin{figure}[h]
	\centering
	\includegraphics[width=\columnwidth]{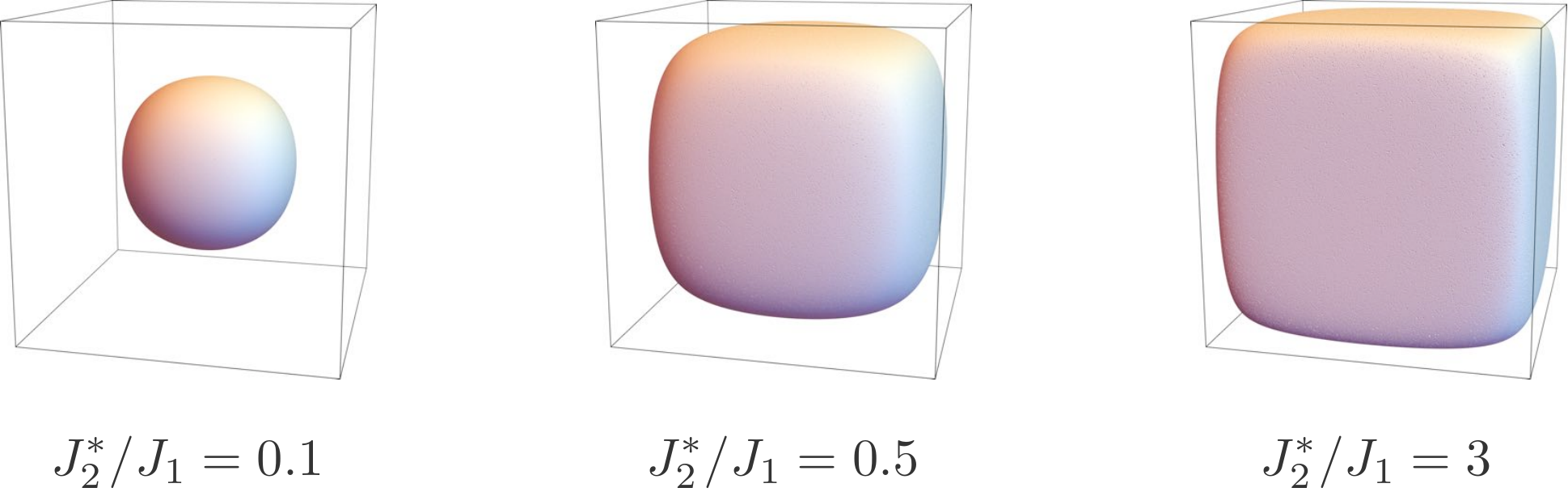}
	\caption{Spin spiral surfaces for the {\bf bcc lattice} 
			$J_1$-$J_2^*$ Heisenberg model for different ratios of the couplings.}
	\label{fig:BCCSpiralSurfaces}
\end{figure}

Another interesting example in three spatial dimensions might be the $J_1$-$J_2$ model on the body-centered cubic (bcc) lattice \cite{Shender1982}.
We note that like the diamond lattice, the bcc lattice is a bipartitice lattice whose two identical sublattices are Bravais lattices
(i.e.~simple cubic lattices). One might therefore hope to find similar spin spiral surfaces as for the diamond lattice when introducing a next-nearest neighbor coupling $J_2$. This is indeed the case, but only if one restricts the next-nearest coupling $J_2$ to be defined by bond distance (which we denote as $J_2^*$ in the following). As illustrated in Fig.~\ref{fig:BCCSpiralSurfaces}, a spherical spin spiral surface forms for $J_2^*/J1 > 1/16$, which deforms into a cube for increasing $J_2^*$ and ultimately matches the boundary of the Brillouin zone in the limit $J_2^* \to \infty$. 

\begin{figure}[t]
	\centering
	\includegraphics[width=\columnwidth]{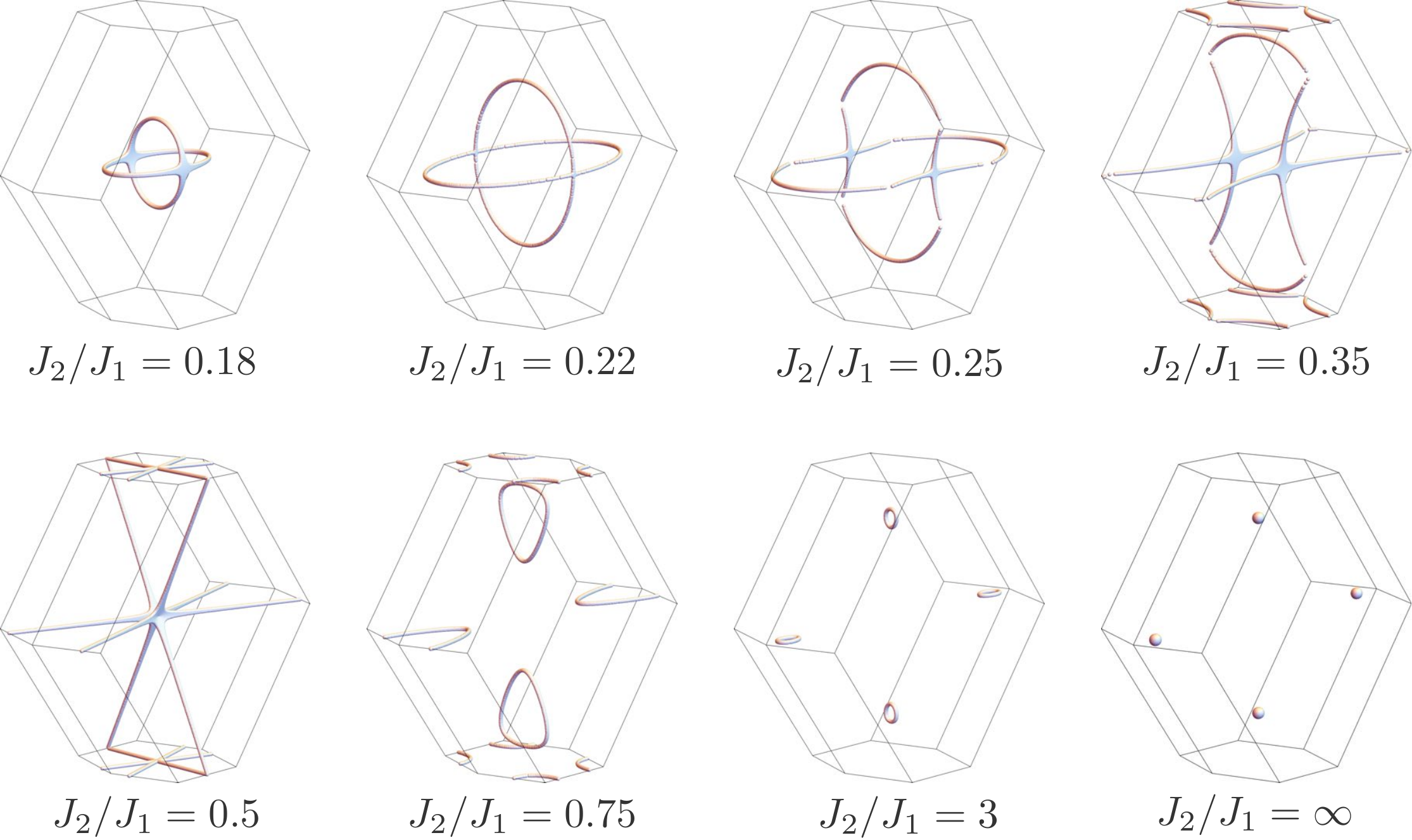}
	\caption{Spin spiral surfaces for the {\bf hyperhoneycomb lattice} 
			$J_1$-$J_2$ Heisenberg model for different ratios of the couplings.}
	\label{fig:HyperhoneycombSpiralSurfaces}
\end{figure}

As a final example we mention the $J_1$-$J_2$ model on the hyperhoneycomb lattice. The hyperhoneycomb is an example of a tricoordinated lattice in three spatial dimensions \cite{Wells1977} and as such closely related to the two-dimensional honeycomb lattice. One might therefore again expect to find spin spiral surfaces upon introducing a next-nearest neighbor coupling $J_2$. However, for this non-Bravais lattice (with four sites in the unit cell), a more subtle picture emerges. While the energy minimization indeed results in degenerate surfaces (illustrated in Fig.~\ref{fig:HyperhoneycombSpiralSurfaces} of the appendix), the Luttinger-Tisza constraint (requiring uniform spin length) is found to be violated for most spiral states on the surface. In fact, the true ground-state degeneracy is found to be described by {\em lines} for all $J_2/J_1 > 1/6$,
as first pointed out in Ref.~\cite{Lee2014}. The evolution of these spiral lines is illustrated in Fig.~\ref{fig:HyperhoneycombSpiralSurfaces} for various coupling strength. Note that only in the limit $J_2 \to \infty$ the spiral lines are found to collapse onto a set of symmetry-related, individual points indicating a single magnetic ground state (without any degeneracies). For an illustration of an individual sublattice see Fig.~\ref{fig:HyperhoneycombLattice} of the appendix.


\subsection{Reverse spin fermion correspondence}

\begin{figure}[t]
	\centering
	\includegraphics[width=\linewidth]{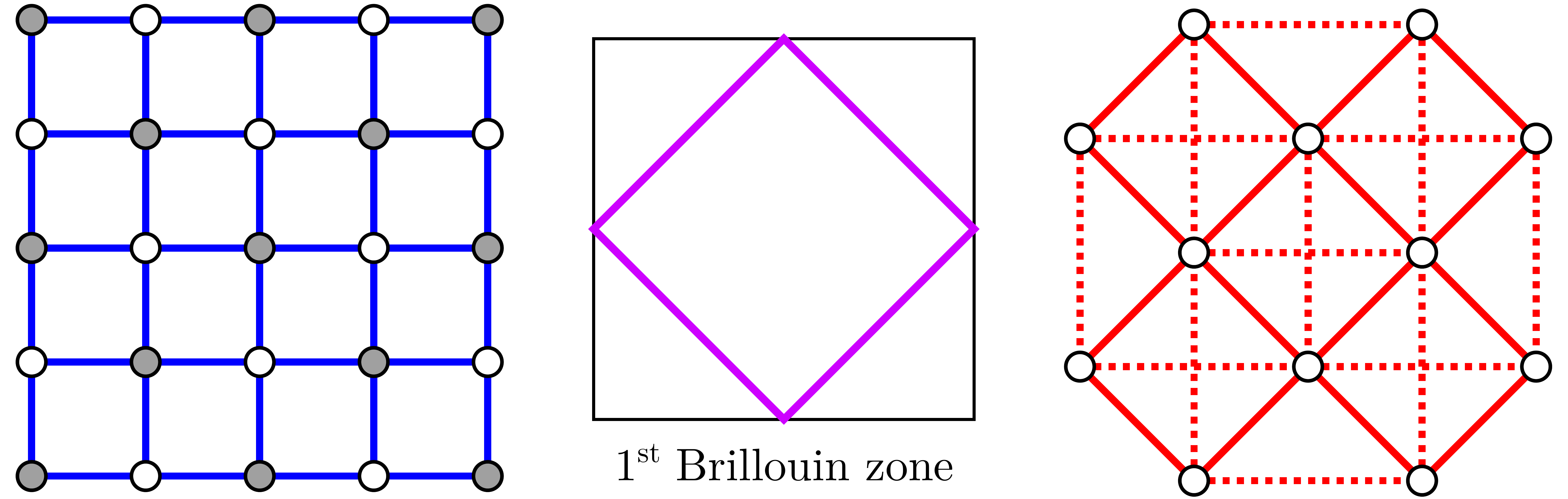}
	\caption{An example of a {\bf reverse spin fermion correspondence}. 
		     Starting from a free fermion hopping model on the square lattice (left), one can construct a frustrated spin model
		     on one of its sublattices which is a fully-connected square lattice as illustrated on the right. 
		     The nearest neighbor coupling $J_{\times}$ is indicated by the solid red lines, 
		     while the next-nearest neighbor coupling $J_{+}$ couples
		     all sites at a {\em bond} distance 2 as illustrated by the dashed red lines.}
	\label{fig:square-lattice-fermion}
\end{figure}

As a final case study we discuss a reverse example of the spin fermion correspondence, i.e.~the explicit construction of a frustrated
spin model from a given fermion hopping model. To do so, we consider free fermions on the square lattice as a starting point. 
Diagonalizing the real-space Hamiltonian via a Fourier transformation, their dispersion is given by
\[
	\varepsilon(\vec{k}) = 2t \left( \cos(k_x) + \cos(k_y) \right) \,,
\]
which readily identifies the nodal structure at the Fermi energy $\varepsilon(\vec{k}) = 0$ as a Fermi {\em line} parametrized by
\begin{equation}
k_x = \pm \left(k_y \pm \pi\right) \,
\end{equation}
as illustrated in the middle panel of Fig.~\ref{fig:square-lattice-fermion}.

To construct a spin model whose spiral ground-state manifold is described by precisely these lines in momentum space,
we again resort to the lattice construction of Section \ref{sec:LatticeConstruction}. The spin model is simply defined by the 
lattice of next-nearest neighbors of the square lattice -- this is a fully connected square lattice illustrated in the right panel
of Fig.~\ref{fig:square-lattice-fermion}. Besides a nearest neighbor exchange indicated by the solid red lines and denoted by
$J_{\times}$ in the following, there is a next-nearest neighbor coupling indicated by the dashed red lines and denoted by 
$J_{+}$ in the following. The full spin Hamiltonian thus reads 
\begin{equation}
	\mathcal{H} = \sum_{\substack{\langle i,j \rangle \\ + \text{bonds}}} J_{+} \vec{S}_i \cdot \vec{S}_j \;
			     +  \sum_{\substack{\langle i,j \rangle \\ \times \text{bonds}}} J_{\times} \vec{S}_i \cdot \vec{S}_j \,.
\end{equation}
To exactly map to the fermion model the two couplings need to be tuned to 
\begin{equation}
J_{\times} = 4t^2 \quad J_{+} = 2t^2 \,.
\end{equation}
For this ratio of coupling strengths $J_{\times} / J_{+} = 2$ will the spin model exhibit a ground-state degeneracy of coplanar spirals described by momenta along the lines captured by the Fermi surface of the fermionic model. Slightly detuning from this coupling ratio will immediately split this degeneracy and result in conventional magnetic order. 


\section{Topological band structures}
\label{sec:topo}

The spin fermion correspondence provides a stringent mapping between the ground states of the Luttinger-Tisza spectrum of the spin interaction matrix and fermionic eigenstates at the Fermi energy. In the case studies above our focus has been to exemplify this correspondence for various spin models where we could map  degenerate spin spiral ground-state manifolds to the Fermi surfaces of 
matching fermion models. Now we want to take the opposite approach and ask whether certain aspects of a given electronic band structure
that go beyond the Fermi surface can reveal themselves also in the corresponding spin models. In particular, we are interested in asking whether topological features such as edge states that can occur in electronic band structures have an equivalent feature in the Luttinger-Tisza spectrum. Going beyond ground states one could also ask whether features further up in the energy spectrum, such as Landau levels, might have any bearing also in the corresponding spin model. 

\begin{figure}[b]
	\centering
	\includegraphics[width=\linewidth]{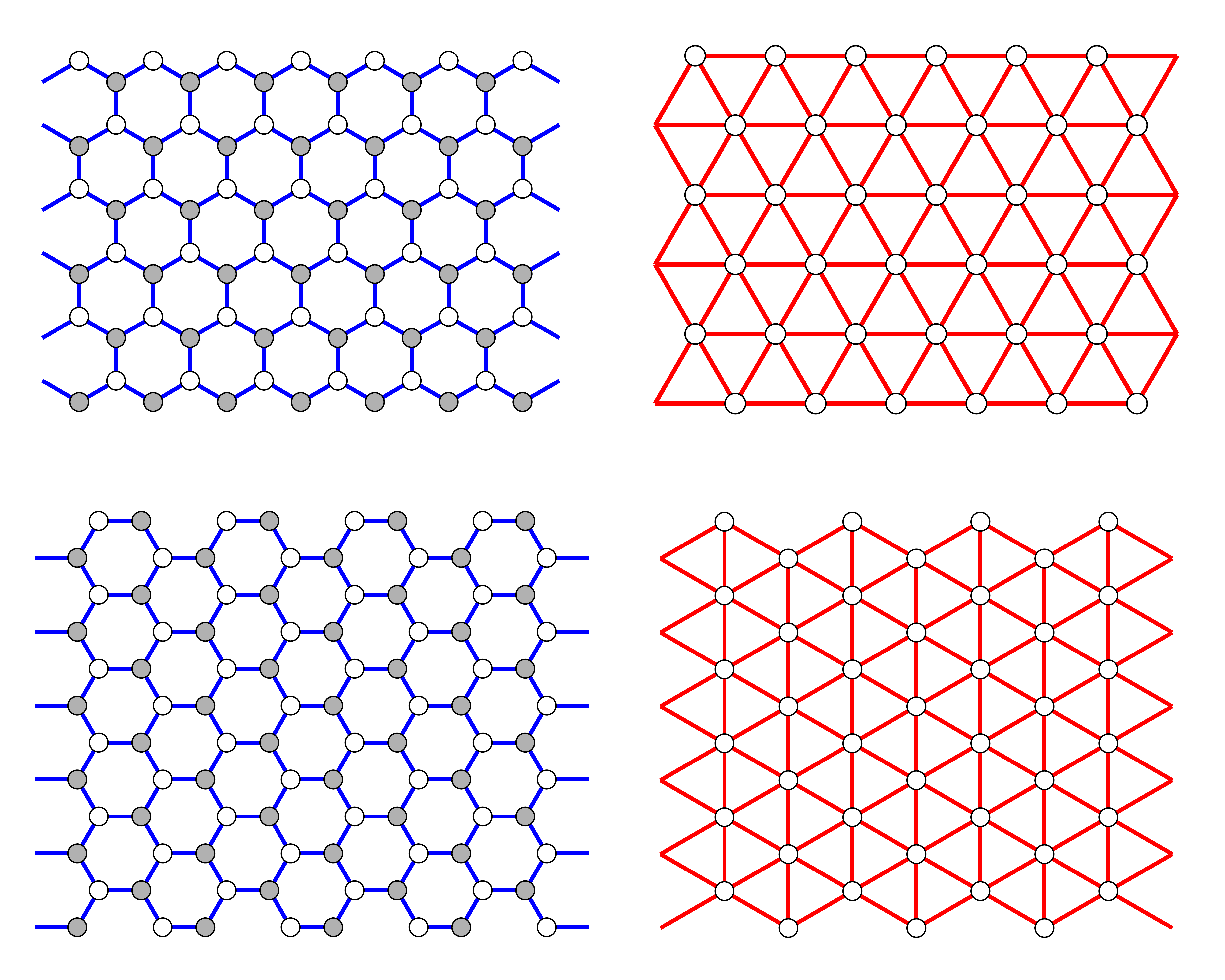}
	\caption{Different {\bf boundary conditions} of the honeycomb lattice (left column) and one of its triangular sublattices (right column).
			The upper row shows zigzag, the lower row shows armchair boundaries along the $x$-direction.}
	\label{fig:graphene-strip}
\end{figure}


\subsection{Edge states}

The occurrence of edge states in a simple tight-binding model is well known from the physics of graphene
-- depending on the realization of boundary of the underlying graphene flake, gapless modes can be localized at the edge of the sample \cite{Nakada1996}.
Zigzag boundaries are found to harbor gapless edge modes, while armchair boundaries do not exhibit any such states \cite{Nakada1996,Tao2011}. 
These experimental observations are in full agreement with the theoretical expectation derived from a simple tight-binding model on the honeycomb lattice realizing these different choices of boundary conditions \cite{Akhmerov2008}.

\begin{figure}[t]
	\centering
	\includegraphics[width=\linewidth]{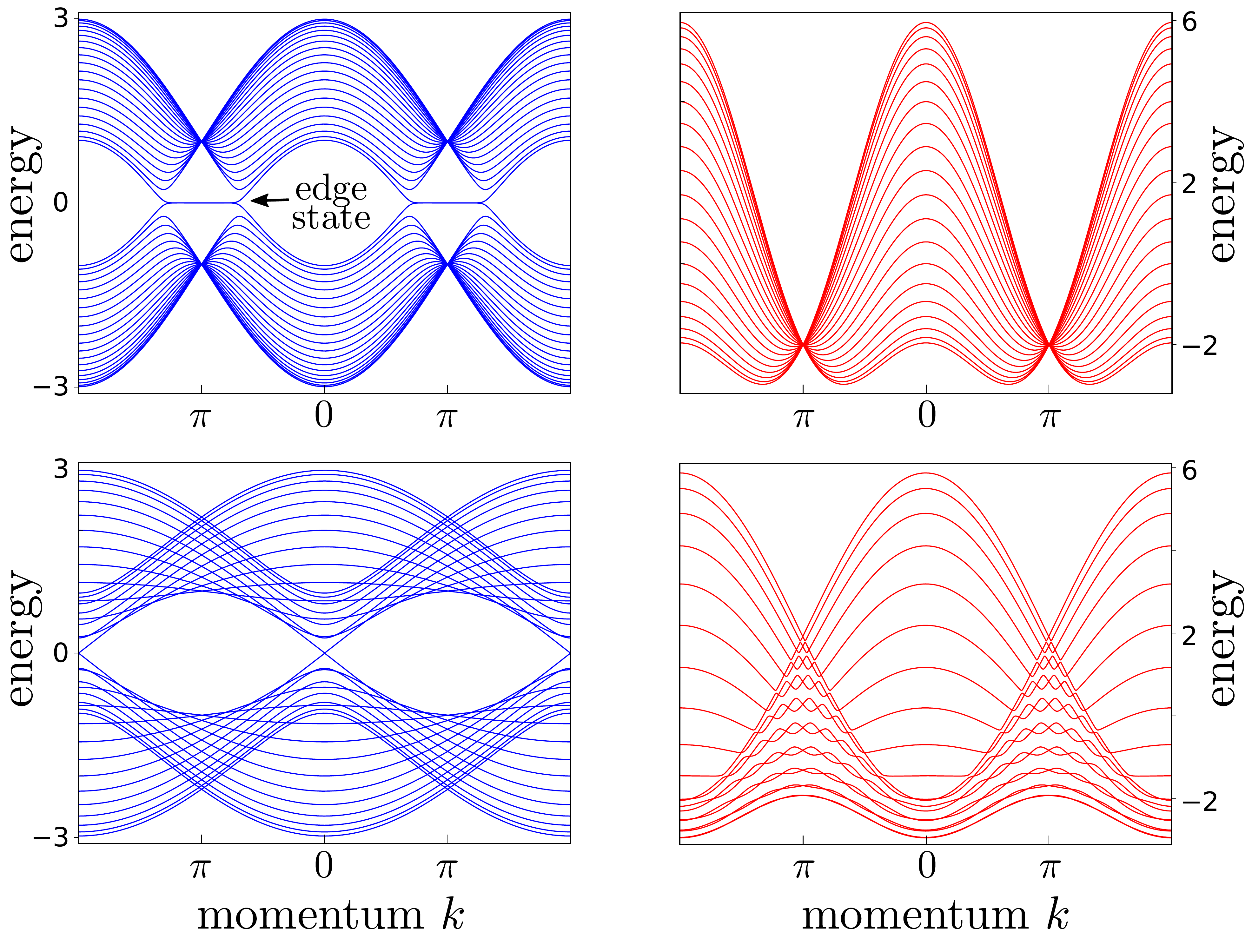}	
	\caption{Occurrence of {\bf edge states} in the band structure of free fermions on the honeycomb lattice 
			and the corresponding antiferromagnet on the triangular lattice. 
			While the fermionic model exhibits an edge state for zigzag boundaries (upper left panel) and
			none for armchair boundaries (lower left panel), the corresponding triangular spin model exhibits
			no edge modes for either boundaries (right column).}
	\label{fig:edge-states-graphene}
\end{figure}

In the context of the spin fermion correspondence at the heart of the current manuscript, this observation motivates the question whether
the 120$^\circ$ order of the triangular antiferromagnet -- the spin analogue of the fermionic honeycomb model -- exhibits gapless edge modes depending on the choice of boundary conditions. As illustrated in Fig.~\ref{fig:graphene-strip} the two different honeycomb boundary conditions map to two distinct boundary conditions also for the corresponding triangular lattice spin model. Upon performing a Luttinger-Tisza calculation for these different boundary conditions the energy spectra of the spin model are found to be in general correspondence to their respective fermionic equivalents, i.e.~the Luttinger-Tisza spectrum of the classical model is the ``square" of the fermionic spectrum as required by the spin fermion correspondence of Eq.~\eqref{eq:square-relation-fermions-squared} -- see the side-by-side comparison of Fig.~\ref{fig:edge-states-graphene}. However, a crucial feature is missing -- the edge state of the fermionic spectrum, manifesting itself as a flat band at the Fermi energy, is strikingly missing from the classical spectrum. In retrospect, this might not come as too much of a surprise since an edge mode is a {\em localized} feature of the quantum system, i.e.~a feature of the ground-state wavefunction that exponentially decays as one moves away from the boundary. For the classical system, however, the Luttinger-Tisza approximation strictly requires that all spins exhibit {\em uniform} length across the system thus inherently excluding the emergence of any spatially localized features. What might be counterintuitive is that
the edge mode is missing from the Luttinger-Tisza spectra of Fig.~\ref{fig:edge-states-graphene} which have been calculated without enforcing the uniform spin length. In fact, it is a different, more algebraic mechanism that prevents the edge states to make an occurrence in the classical spectrum.

Recall that the spin fermion correspondence requires that the classical interaction matrix ${\bf M}$ is not simply the squared fermionic hopping matrix $\mathbf{H}^2$, but that there is an additional diagonal matrix $\mathbf{E}^*$ of the form
\[
\mathbf{M} = \mathbf{H}^2 - \mathbf{E}^* \,.
\]
Upon close inspection, one finds that the elements of $\mathbf{E}^*$ are proportional to the number of nearest neighbors of the respective sites as required by Eq.~\eqref{eq:GoundStateEnergy}.
For a system with periodic boundary conditions the matrix $\mathbf{E}^*$ is therefore proportional to the identity matrix and results in a simple shift of the classical spectrum. However, for systems with open boundary conditions as considered here this is no longer the case. With the sites at the boundary of the system having a smaller number of neighbor sites than those in the bulk, one finds that the first and last element of the diagonal of $\mathbf{E}^*$ are in fact smaller than all other diagonal elements. It is exactly this effect that algebraically eliminates the edge states from the classical spectrum.


\subsection{Landau levels}

\begin{figure}[t]
	\centering
	\includegraphics[width=0.47\linewidth]{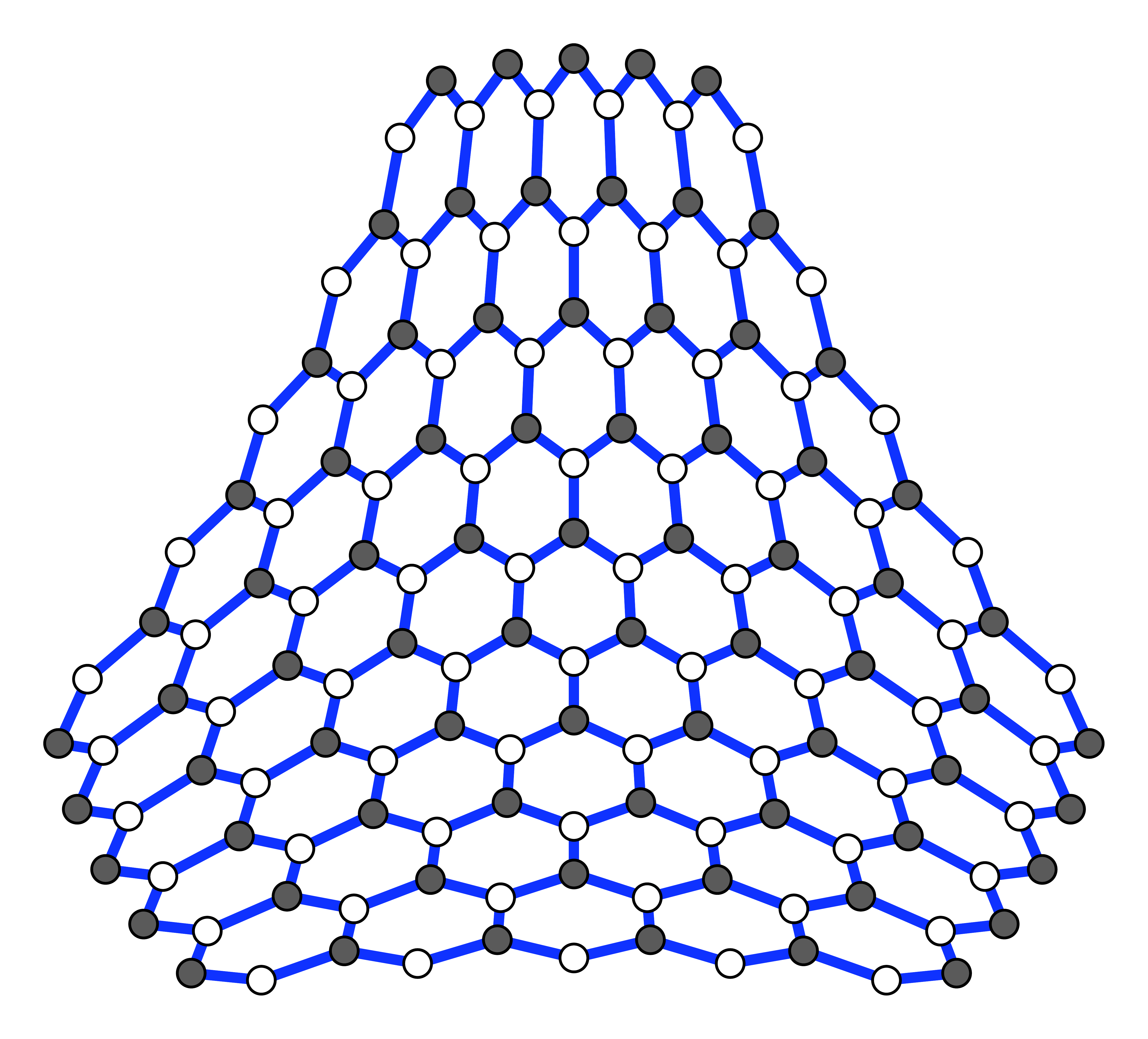}	
	\hspace{0.04\linewidth}
	\includegraphics[width=0.47\linewidth]{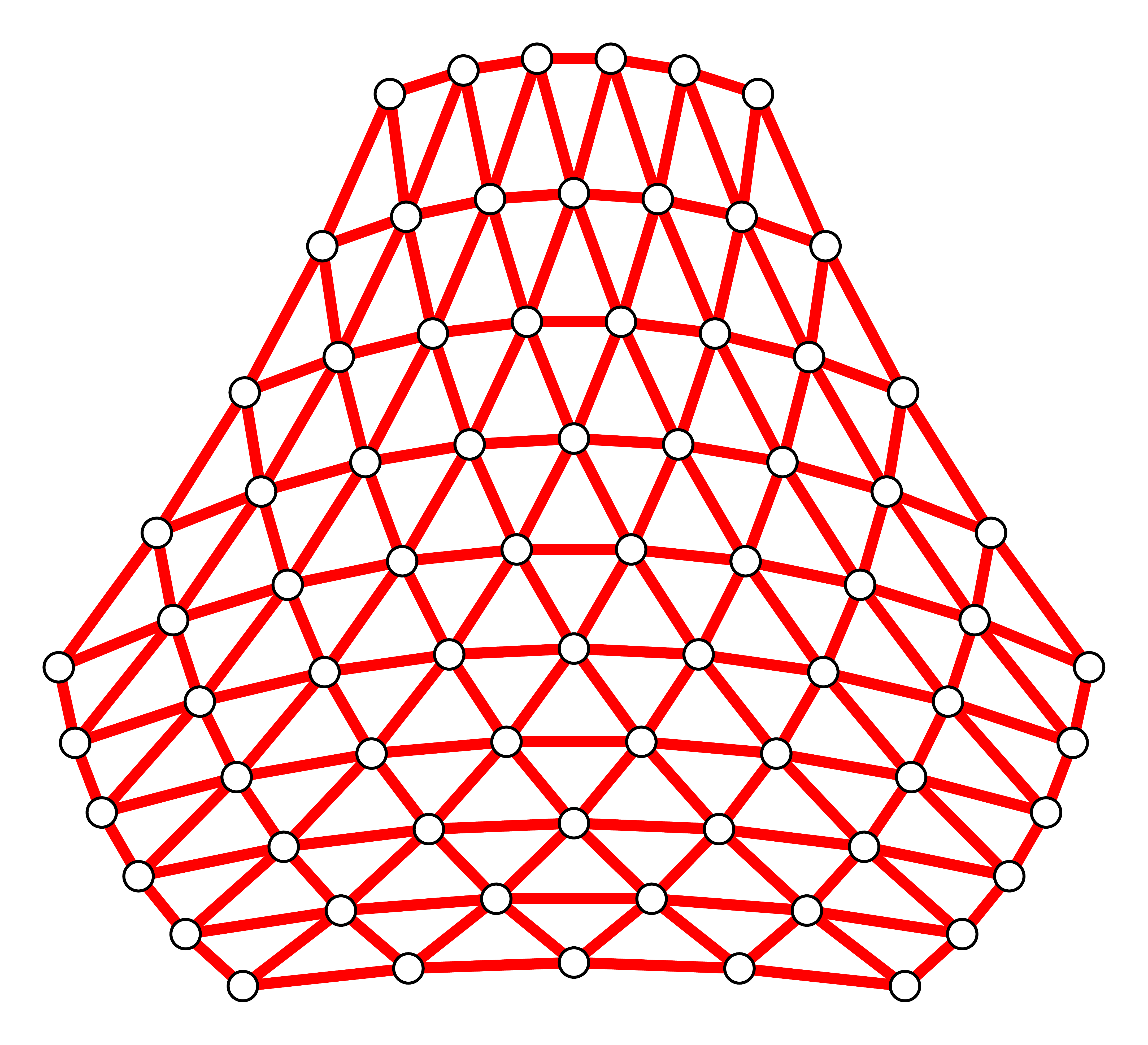}	
	\caption{Deformation of the honeycomb and triangular lattice due to {\bf triaxial strain}.
		}
	\label{fig:graphene-flakes}
\end{figure}

Another idea that one might want to entertain is the question whether the spin fermion correspondence allows for an analogue of Landau level physics in a frustrated antiferromagnet. While the emergence of Landau levels is typically associated with the application of an external magnetic field, an alternative route of inducing Landau levels has been explored via the application of {\em triaxial strain} (inducing a pseudomagnetic field in form of an artifical gauge-field vector potential)  in the context of graphene \cite{Guinea2010}. 
Again, this physics can be captured by a simple honeycomb tight-binding model -- in this case with spatially modulated hopping amplitudes reflecting the effect of the triaxial strain, which is visualized in the left panel of Fig.~\ref{fig:graphene-flakes}. One might therefore ask what the effect of triaxial strain and a similarly spatially modulated coupling strength is on the triangular lattice antiferromagnet, see the right panel of Fig.~\ref{fig:graphene-flakes}.
By virtue of the spin fermion correspondence, one indeed finds that the spectrum of the interaction matrix of the strained triangular lattice antiferromagnet exhibits a discrete level spacing above the ground-state energy (not shown). However, all of these states do not reflect the actual physical states of the classical antiferromagnet as they generically do not obey the uniform spin length requirement. 

Nevertheless, it might be an interesting direction for future exploration to ask whether triaxially strained triangular antiferromagnets exhibit any topological features in their ground states.

\subsection{Topological insulators}

Considering topological features of fermionic band structures, one might further ask whether the topological insulator
has a classical spin analogue that can be motivated via the spin fermion correspondence. While this is a tempting idea
that would bear some similarity to the concept of topological mechanics introduced by Kane and Lubensky \cite{KaneLubensky2014}, 
such a scenario is prevented for several reasons. 
First, going back to the general form of the fermionic Hamiltonian arising
from the spin fermion correspondence \eqref{eq:FermionMatrixBlocks} discussed in Section \ref{sec:SymmetryClass} 
one finds that the fermionic system naturally resides in symmetry class BDI. From the general classification of topological
insulators \cite{Schnyder2008,Kitaev2009,Ryu2010} rooted in the 10-fold way symmetry class classification it is known 
that symmetry class BDI does {\em not} allow for the occurrence of a topological insulator in two and three spatial dimensions, 
which are of interest here, but only in one spatial dimension.
Second, we note that even if the fermionic band structure would have allowed for the occurrence of a topological insulator one
of its key features -- protected gapless edge states on its surface -- would not be possible to realize in the spin model (as discussed 
above).


\section{Discussion}
\label{sec:discussion}

The spin fermion correspondence derived in this manuscript relates the spin spiral surface of frustrated antiferromagnets
with the Fermi surface of non-interacting fermion systems. Despite this close relation one should note that there exist some
fundamental differences between the two manifolds. One such distinction arises when considering the effect of small perturbations.
A Fermi surface is a remarkably stable object in that most perturbations (but pairing or nesting instabilities) leave it mostly untouched
and merely deform its shape. In contrast, the spin spiral surface is an exceedingly unstable object. Any residual interaction added 
to the original Heisenberg Hamiltonian typically destroys the degenerate spin spiral manifold and favors some sort of spin spiral order. 
For some systems the spin spiral surface 
is unstable even to entropic effects such as an order-by-disorder transition \cite{Villain1980} at finite temperatures. 
As such the spin spiral manifold of a classical antiferromagnet oftentimes governs only its ground state at zero-temperature. 
In fermionic systems, on the other hand, knowledge of the Fermi surface immediately allows to infer some thermodynamic signatures  such as the
leading contribution to the specific heat at low temperatures (a power law dominated by the co-dimension of the Fermi surface).

An interesting aspect of the spin fermion correspondence is that it points a way to inferring the ground-state physics of a frustrated magnet (which might be hard to access) from the band topology of a free-fermion system at the Fermi energy (which might be readily available) and vice versa. 
In practice, the most intriguing example of such a transfer would have been to find topological aspects of the fermion system reincarnate themselves in the spin system. While we have explored this idea in the context of edge states to no success, it remains to be seen whether it works in other instances such as the suggestion that the ground state of the triaxially strained triangular lattice antiferromagnet might exhibit topological features similar to the corresponding triaxially strained fermion model on the honeycomb lattice. The reverse direction -- inferring features of the fermionic band structure from knowledge about the ground state of the corresponding spin model -- has proved insightful in constructing simple fermion models that exhibit completely flat bands at the Fermi energy as demonstrated for the extended honeycomb and diamond lattices. This might be a useful starting point for the further construction of fractional Chern insulators or other non-trivial states that are generated from an interaction induced splitting of such highly-degenerate flat bands.

Taking a step back, we note that the spin fermion correspondence introduced in this manuscript is in some way orthogonal to the concept 
of {\em fractionalization}, which is often employed in the context of {\em quantum} spin liquids. The latter are conceptualized as 
macroscopically entangled quantum states where the original spins decompose into novel, emergent degrees of freedom that 
carry a fractional quantum number \textendash\  partons and a gauge field. Oftentimes, the emergent parton is a {\em fermionic}
degree of freedom and the spin liquid state is understood as a metal formed by these emergent fermions.
Familiar examples of this fractionalization include the representation of the original spins in terms of Majorana  fermions 
(coupled to a $Z_2$ gauge field) or Abrikosov fermions (coupled to a $U(1)$ gauge field).
The former scenario is well known from the analytical solution of the Kitaev model in two and three spatial dimensions \cite{Kitaev2006,OBrien2016}, where the emergent Majorana fermions form various types of Majorana metals with Fermi surfaces,
nodal lines, or Weyl/Dirac points depending on the geometry of the underlying lattice 
\cite{Kitaev2006,Mandal2009,Hermanns2014,Hermanns2015,OBrien2016}.
The latter scenario, a decomposition of the original spins into Abrikosov fermions and a $U(1)$ gauge field, has been proposed, for instance, for the kagome antiferromagnet \cite{Hastings2000} where the emergent
Abrikosov fermions form a Dirac semi-metal. This picture of fractionalization, which also makes a correspondence between
a spin system and a fermion metal, should be contrasted to the spin fermion correspondence described in the manuscript at hand.
Here we note that in order to arrive at the fermion model it takes {\em two} classical spin systems (each constituting one sublattice 
of the fermion lattice model), i.e. in a certain sense one ``doubles" the classical system to arrive at the fermion system and thereby
reverses the idea of fractionalization.

Going beyond spin models, we note that the classical to quantum correspondence of this manuscript establishes an exact
connection between a {\em minimization problem} (such as the identification of the ground-state manifold of a classical spin system) 
and the widely studied physics of Fermi surfaces. 
This is an intriguing avenue for further exploration as more general minimization problems 
might in fact fit the classical to quantum correspondence more generically than the spin fermion case study at hand where 
the additional Luttinger-Tisza constraint of uniform spin length has obscured some aspects of the general correspondence.
One such more general example occurs in the context of Ginzburg-Landau theory where the free energy can be expressed in terms
of a complex, momentum-dependent order parameter $\Phi_{\vec{k}}$ by 
\[
	F[\Phi] = \int d^3k \;\; \Phi^*_{\vec{k}} \left( r\cdot {\bf 1} + {\bf M}(\vec{k}) \right) \Phi_{\vec{k}} \;+ \;U \Phi^4 \; + \ldots \,,
\]
where $r$ is some control parameter tuning the system through the phase transition (such as $r = T-T_c$ for a thermal phase transition)
and ${\bf M}(\vec{k})$ is a general, momentum-dependent interaction matrix. To identify the location of the phase transition one needs to
track the minimum (and sign change) of the quadratic term in the Ginzburg-Landau expansion  
-- a minimization problem that via the classical to quantum
correspondence can again be recast in terms of a fermionic system (but without any additional constraints on the order parameter).


\begin{acknowledgments}
We acknowledge discussions with D. Bagrets, A. Ludwig,  A. Rosch, and M. Yamada.
This work was partially supported by the DFG within the CRC 1238 (project C02).
The numerical simulations were performed on the CHEOPS cluster at RRZK Cologne.
J.A.  thanks the Bonn-Cologne Graduate School of Physics and Astronomy (BCGS)
and the German Academic Scholarship Foundation (Studienstiftung des deutschen Volkes) 
for support.
\end{acknowledgments}


\appendix

\section{Validity of Luttinger-Tisza calculation}

In contrast to simple Bravais lattices, applying the Luttinger-Tisza approach to non-Bravais lattices (with multiple sites in the unit cell)
requires an additional step beyond the minimization of eigenvalues of the interaction matrix. One needs to carefully analyze that the
minimal eigenvectors fulfill the Luttinger Tisza ``hard spin" constraint. In order to provide a meaningful description of a ground-state
spin configuration for an $O(3)$ Heisenberg model, every minimal eigenvector $\tilde{S}_{\vec{q},j}$ needs to exhibit a uniform 
spin length for all sites in the unit cell, i.e. it needs to fulfill the condition $|\tilde{S}_{\vec{q},j}^A|^2 = 1$ for all components $A$
(corresponding to the different sites in the unit cell).

There are some special cases in which the constraint is satisfied naturally. First of all, any Bravais lattice satisfies the constraint by construction, as it only has one site per unit cell and therefore any eigenvector only has one component. Examples include the triangular or square lattice.
Moreover, any {\em bipartite} lattice which can be decomposed into two Bravais lattices also generically satisfies the Luttinger-Tisza constraint. 
This can be seen by squaring the interaction matrix which preserves the eigenvectors but results in a block structure for the two sublattices. Examples for this case are the diamond, bcc, and honeycomb lattices.


\subsection*{Luttinger-Tisza calculation for the hyperhoneycomb $J_1$-$J_2$ model}

\begin{figure}[b]
	\centering
	\includegraphics[width=0.45\columnwidth]{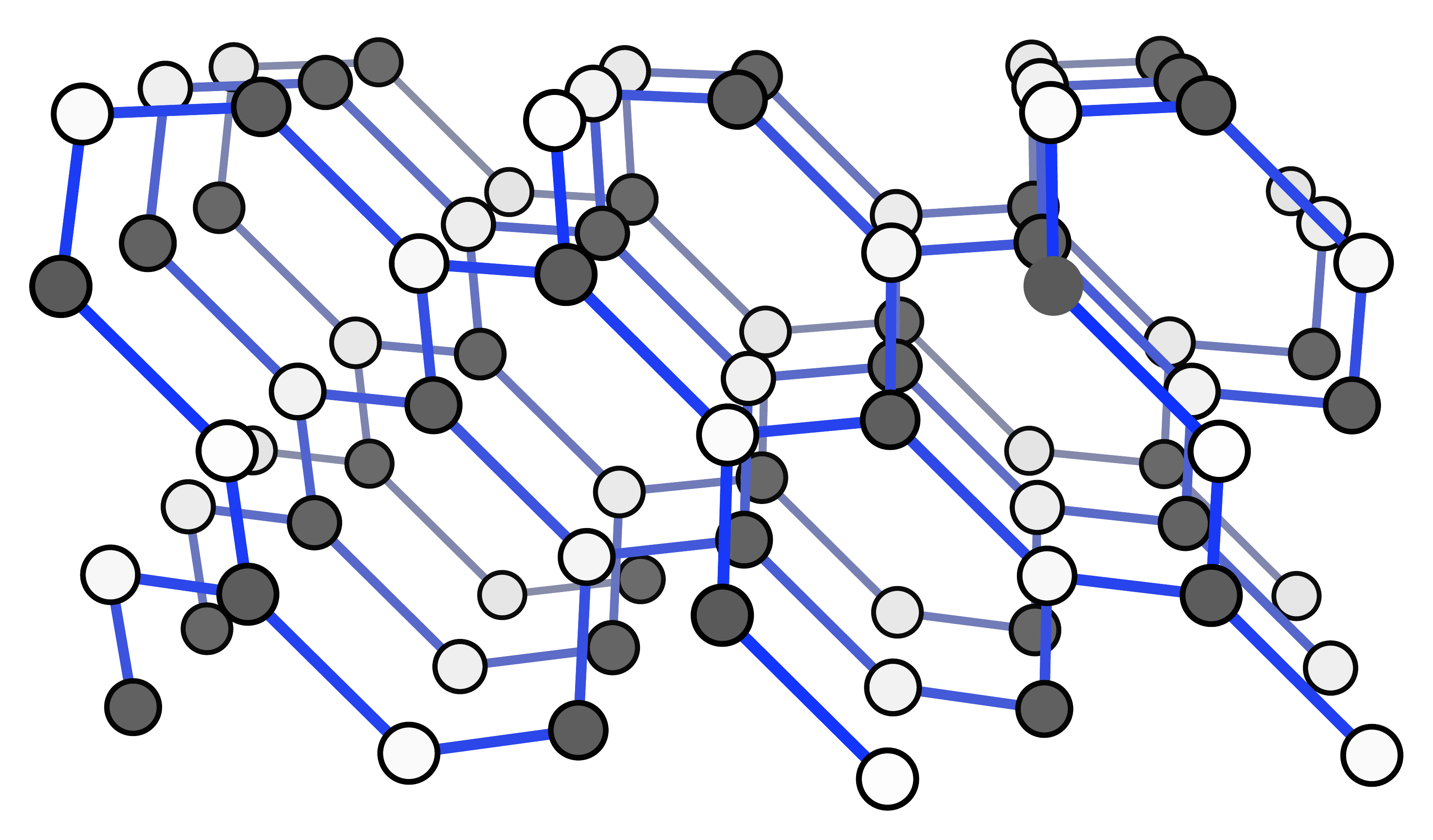}
	\hspace{0.08\columnwidth}
	\includegraphics[width=0.45\columnwidth]{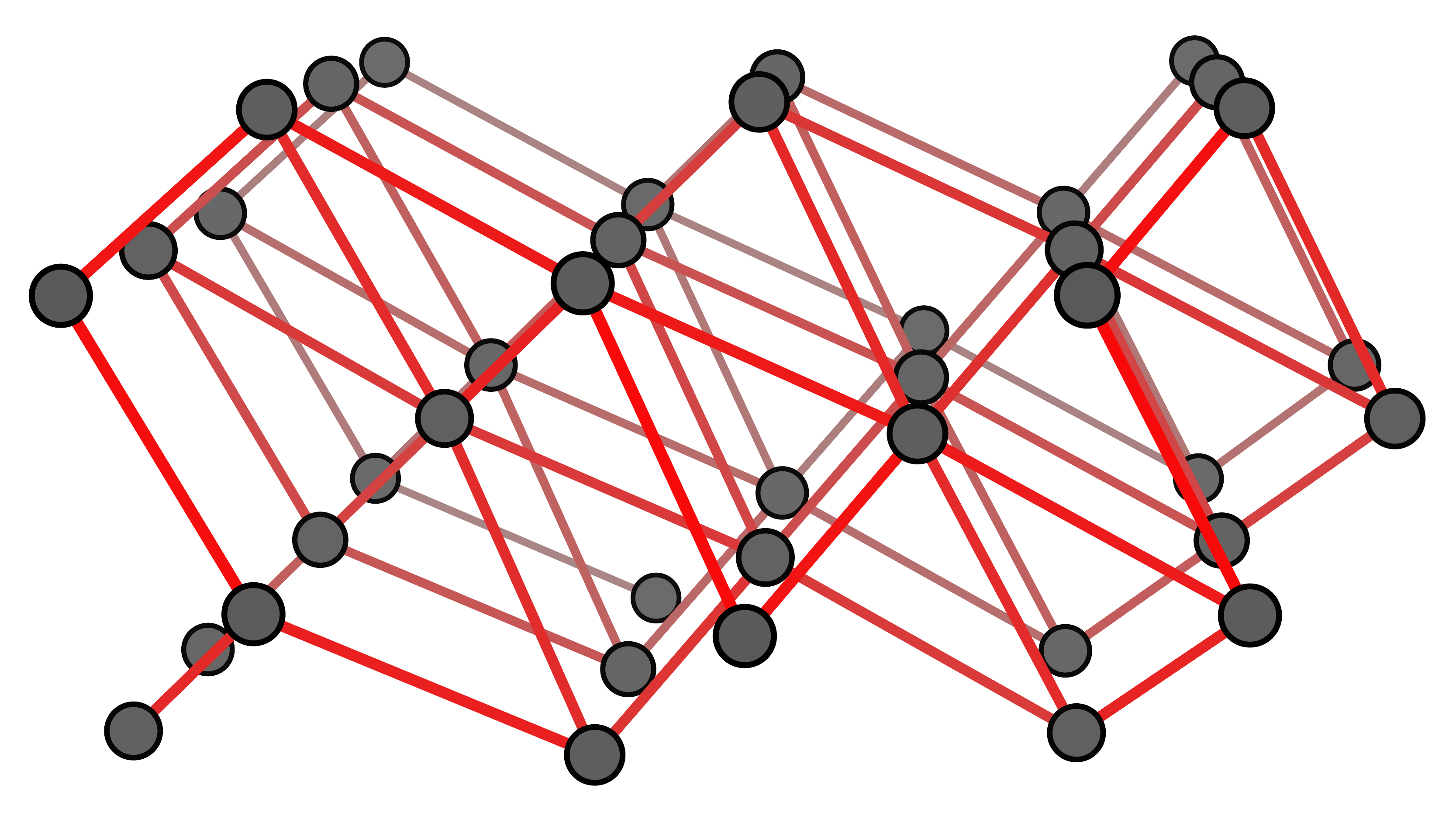}
	\caption{The {\bf hyperhoneycomb lattice} (left) and one of its sublattices (right).}
	\label{fig:HyperhoneycombLattice}
\end{figure}

\begin{figure}[t]
	\centering
	\includegraphics[width=\columnwidth]{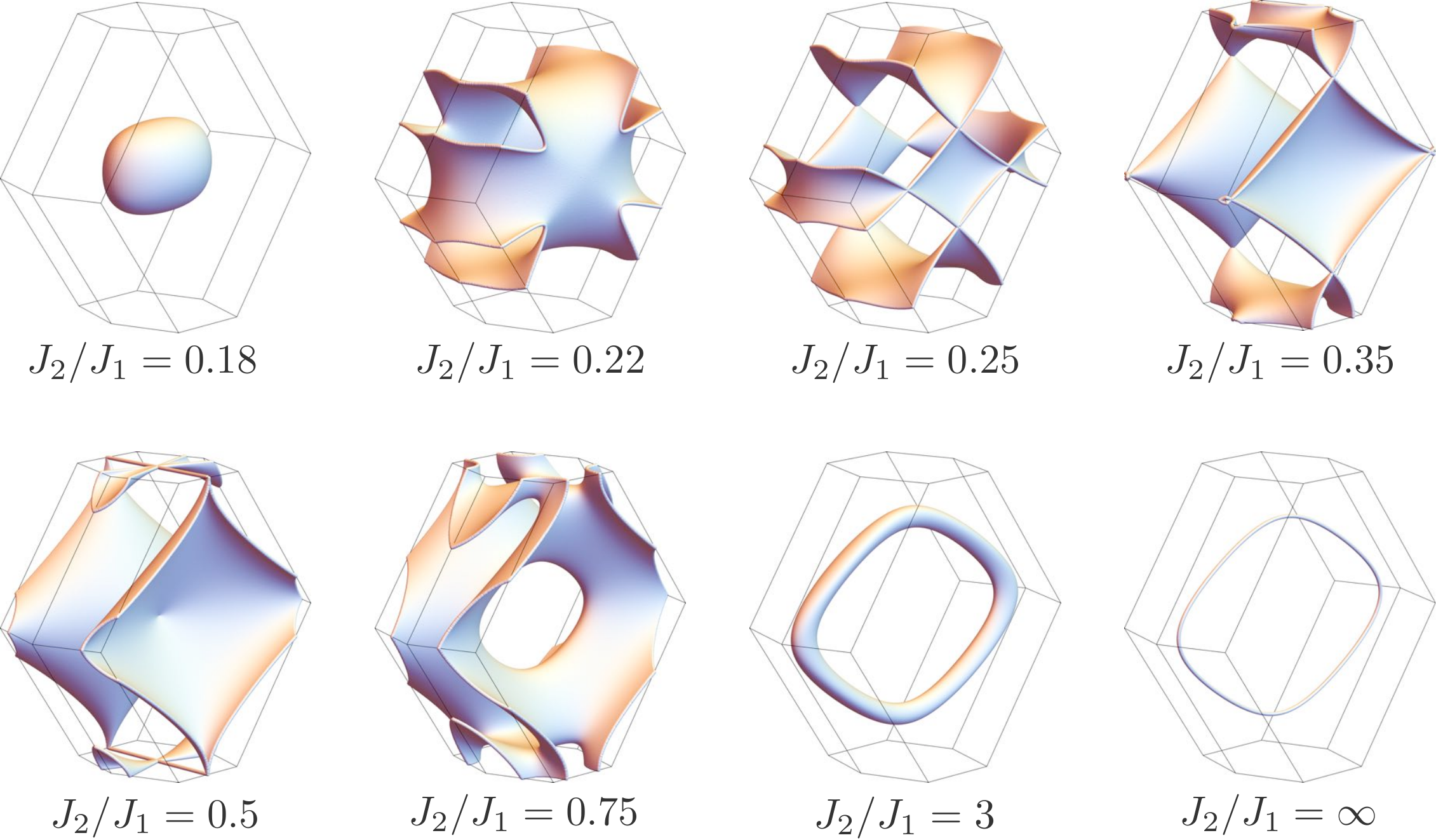}
	\caption{{\bf Minimal energy manifold} after diagonalization of the the spin interaction matrix
			of the $J_1$-$J_2$ Heisenberg model on the hyperhoneycomb lattice
			for various ratios of the couplings $J_2 / J_1$. 
			Note that most states on these manifolds do not fulfill the Luttinger-Tisza contraint of uniform spin length,
			see for comparison Fig.~\ref{fig:HyperhoneycombSpiralSurfaces} which shows the reduced line-like manifolds of these valid ground states.
			}
	\label{fig:HyperhoneycombMinimalEnergySurfaces}
\end{figure}

We close with an example calculation where the diagonalization of the spin interaction matrix results in a large manifold of states,
out of which only a small subset fulfills the Luttinger-Tisza constraint of uniform spin length. The example at hand is the $J_1$-$J_2$ model
on the hyperhoneycomb lattice, an elementary tricoordinated lattice in three spatial dimensions \cite{Wells1977} 
illustrated (together with one of its sublattices) in Fig.~\ref{fig:HyperhoneycombLattice}. 
Diagonalizing the spin interaction matrix for various couplings $J_2$ one generically finds a manifold of states that constitutes a 
{\em surface} in the Brillouin zone as illustrated in Fig.~\ref{fig:HyperhoneycombMinimalEnergySurfaces}. However, most of the states on this
surface are found to violate the uniform spin length constraint and are therefore not valid ground states of the original spin model.
In fact, the subset of valid ground states is reduced to {\em lines} in the Brillouin zone as illustrated in Fig.~\ref{fig:HyperhoneycombSpiralSurfaces} in the main part of this manuscript.
Considering the corresponding fermion model, which can be defined either on a bilayer hyperhoneycomb or, alternatively, as a spinful hyperhoneycomb model, we note that its Fermi surface is given by the {\em full} Luttinger-Tisza minimal energy manifold of Fig.~\ref{fig:HyperhoneycombMinimalEnergySurfaces}, as the fermionic system is not subject to any further constraints.

\bibliography{spirals}

\end{document}